\documentclass[aps,pre,showpacs,floatfix,twocolumn,amsmath,amssymb,dvips]{revtex4-1}

\usepackage{epsfig}
\usepackage{graphicx}
\usepackage{dcolumn}
\usepackage{bm}
\usepackage[colorlinks=true,dvipdfm]{hyperref}

\begin{document}
\title{Renormalization-group theory of dynamics of first-order phase transitions in a field-driven scalar model}

\author{Fan Zhong}
\affiliation{State Key Laboratory of Optoelectronic Materials and
Technologies, School of Physics and Engineering, Sun Yat-sen
University, Guangzhou 510275, People's Republic of China}

\date{\today}

\begin{abstract}
We show by a detailed study of the mean-field approximation, the Gaussian approximation, the perturbation expansion, and the field-theoretic renormalization-group analysis of a $\varphi^{3}$ theory that its instability fixed points with their associated instability exponents are quite probably relevant to the scaling and universality behavior exhibited by the first-order phase transitions in a field-driven scalar $\phi^4$ model below its critical temperature near their instability points. Finite-time scaling and leading corrections to scaling are considered. We also show that the instability exponents of the first-order phase transitions equal those of the Yang--Lee edge singularity and employ the latter to improve our estimates of the former. The outcomes agree well with existent numerical results.
\end{abstract}

\maketitle

\section{\label{intro}Introduction}

Phase transitions are of great importance in almost all
scientific fields. Traditionally, a phase transition is said to be
$n$th order if the $n$th derivative of the chemical potentials
of the two phases involved is discontinuous at the transition
point and all lower-order derivatives are continuous
\cite{Ehrenfest}. Transitions of orders higher than one often show, however,
singularity in their transition points, and nowadays one usually
distinguishes only first-order or discontinuous phase transitions
(FOPTs) from continuous ones that include all the others \cite{fisherc}. The first recorded continuous phase transition was the critical point of CO$_2$
discovered in 1869 by Andrews \cite{Andrews}. The
renormalization-group (RG) theory for it and other continuous phase transitions in general \cite{Wilson} was
developed, however, more than one century later based on
scaling and universality in the critical phenomena exhibit near such transitions \cite{Stanley}.
Some of its results have now been predicted in high precisions
and are consistent with high-quality microgravity experiments
\cite{Barmatz}.

FOPTs, on the other hand, have apparently a far longer history and
appear far more frequently \cite{Gunton83,Binder87,Debenedetti}. Take H$_2$O as an example, its usual phase diagram contains three phases: ice, water, and vapor, which
are common to daily life. All transitions between these phases are FOPTs
except the isolated critical point at the end of the line of
vaporization. The first well-known theory of phase
transitions can be traced back to van der Waals' equation of state
\cite{Waals}, which, besides the description
of the critical point, when combined with Maxwell's equal-area
construction \cite{Maxwell}, predicts that there is a gas--liquid
coexistence region, within which metastable states are separated
from unstable states by a well-defined spinodal curve at which the
isothermal compressibility diverges. Gibbs' theory of equilibrium
\cite{Gibbs} further identifies two distinct mechanisms of
equilibration for the two kinds of nonequilibrium states, viz.,
nucleation and growth \cite{gunton,Oxtoby,Sear} versus unstable
growth (or spinodal decomposition \cite{Binder01} that is often
referred to binary systems), respectively. This picture of the
FOPTs dominates their study for more than one century thereafter.
Nevertheless, a theory of nucleation is still considered to be quite
good at present even if its predicted nucleation rate agrees with that measures only to
within several orders of magnitude \cite{Oxtoby,Sear}; and the
nonlinear theory of spinodal decomposition \cite{LBM} has yet to
be substantially improved \cite{Binder01}. The only modification
to the picture is that the crossover of the two modes of
transition is believed to be smooth rather than sharp and thus neither is
the spinodal curve sharp if exists \cite{Binder,zhongjcp}.
In addition, metastable and unstable states are often accompanied with hysteresis, a non-equilibrium and nonlinear phenomenon that is difficult to control.
Accordingly, a general theory that is comparable to the
well-developed framework of the RG theory for the continuous
transitions has yet to be developed for the FOPTs.

A challenge question along this line is, instead of case study, whether there exist also
scaling and universality in FOPTs that may be used as general
characteristics to study them. In the case of equilibrium,
within the framework of the RG theory, the discontinuity of extensive
variables across an FOPT has been argued to correspond
to the existence of a discontinuity fixed point with at least an eigenvalue
equal to the dimensionality of the system \cite{niehuis}. Trivial
scaling behavior then follows \cite{fisher}. In the dynamic
perspective, there are two main occasions that scaling and
universality emerge. One is usually referred to as phase-ordering
kinetics in which a system evolves to an equilibrium thermodynamic
state which consists of two coexisting phases from a
non-equilibrium metastable or unstable one-phase state which
originates usually from rapidly quenching of a one-phase, thermal
equilibrium state \cite{Gunton83,bray94}. In such systems, one finds that
in the late stages of growth, the structure function scales by a
time-dependent characteristic length scale which characterizes the
size of the growing domains \cite{Marro79}. Several universality classes have
also been established in relation to Model A, B, etc. that are
originally defined in critical dynamics \cite{hohenberg}.
Naturally, the RG theory has been applied to such phase ordering
kinetics to understand the origin of such dynamic scaling \cite{bray94}. However,
lack of a small parameter analogous to the case in the critical phenomena
renders such approaches essentially a scaling analysis
\cite{bray94}. As many practical transitions are driven by an
external field or the temperature in such a way that the
systems change from one phase completely to the other, it has been
found that energy dissipations and/or hysteresis exhibit scaling
with respect to the sweep rate of field which serves as a
characteristic of the irreversible processes involved
\cite{zhang86,Rao,zhong94,zhonge1,zhonge2,zhang95,zhongssc,cha}. An RG theory has also been adapted successfully to the dynamic scaling of hysteresis in a toy vector model with
infinite number of vector components and subjecting to a varying external
field, arriving at scaling forms with respect to the
field sweep rate that are determined by a zero-temperature fixed
point and agree excellently with numerical results \cite{zhongl}.
A subsequent attempt to a more practical Ising model using Monte
Carlo RG method has not yet been able to determine the fixed point
unambiguously \cite{zhongr}. Recently, focusing on the dynamics of
a generic FOPT driven by an external field in the $\phi^{4}$ model
below its usual critical point, we have shown by a field-theoretic RG
method that it is governed by an unexpected instability fixed point
of a corresponding $\varphi^{3}$ model. Accordingly, it does
exhibit a distinct scaling and universality behavior with
corresponding instability exponents different from the critical ones
\cite{zhongl05}. Here we shall follow a more standard approach and study in detail a mean-field theory, a Gaussian theory, a perturbation expansion about the mean-field theory, and a field-theoretic RG theory of the $\varphi^3$ theory. We also employ the Yang--Lee edge singularity to improve the estimates of the instability exponents. The good agreement of these estimates with existent numerical results, together with the mean-field numerical and analytical outcomes, shows that, although in the mean-field theory, the FOPT and the third-order transition described by the $\varphi^3$ theory fall into opposite ground states and appear different, they are governed most likely by the same instability point and hence instability fixed points. We also discuss how the imaginary fixed points are reached, which is related to the relevancy of the fixed points.

The rest of the paper is organized as follows. Starting with a
usual scalar $\phi^4$ model below its critical temperature
in the presence of an external field, we derive an associated
$\varphi^3$ model that is relevant to the FOPTs involved
in Sec.~\ref{model}. Dynamics of the model is also defined there. After a brief review of the dynamic field theory in Sec.~\ref{dft} to set the stage, we then present details of the mean-field theory in Sec.~\ref{mft}, followed by the Gaussian theory (Sec.~\ref{gauss}), which describes fluctuations around the mean-field theory and shows clearly a divergent correlation length and a divergent correlation time similar to the critical phenomena, and the one-loop perturbation expansions (Sec.~\ref{pert}), which exhibits infrared divergences below the upper critical dimension of 6. Detailed exposition of the RG theory to deal with the divergences appears in Sec.~\ref{rgt}. We then argue in Sec.~\ref{YL} that the large-scale behavior of the scalar $\varphi^3$ theory for FOPTs falls in the same universality class as the Yang--Lee edge singularity \cite{Fisher78} and employ its existent exponents to two and three loop approximations to estimate the instability exponents for the FOPTs. As the infrared stable fixed point found for the $\varphi^3$ theory is purely imaginary numbers, we enter into discussions in Sec.~\ref{discuss} on concern about whether the fixed point describes true asymptotic scaling behavior or just crossover. A summary is given in Sec.~\ref{sum}. Two appendices are included that sketch briefly the formulation of a supersymmetry dynamic action (Appendix~\ref{sa}) and the computations of the relevant integrals and expansions (Appendix~\ref{use}).

\section{\label{model}Model}
\subsection{\label{phi4}Scalar $\phi^4$ model}
We consider a model with a usual Ginzburg-Landau functional
\begin{equation}
{\cal H}[\phi ] = \int {d{\rm {\bf x}}\left\{ {\frac{1}{2}r\phi ^2
+ \frac{1}{4!}g\phi ^4 + \frac{1}{2}[\nabla \phi]^2 - H\phi }
\right\}} \label{H}
\end{equation}
of a scalar order parameter $\phi$ in the presence of an external field $H$, where $g$ is a coupling constant and is positive for stability and $r=c_1(T-T_c)$ is the reduced temperature
with $T_c$ being the mean-field critical temperature and $c_1$ a positive constant. The total free energy is
\begin{equation}
{\cal F}=-\ln Z = -\ln\int{\cal D}\phi\exp\{-{\cal H}[\phi]\},
\label{f}
\end{equation}
where the functional integral is over all possible configurations and we have absorbed the temperature factor into the definition of ${\cal H}$.
The free energy thus obtained is the true one in the sense that it describes
equilibrium properties of the system concerned. Therefore, it ought to
be a convex function of the order parameter and thus possesses no
metastable and unstable states. Analytical continuation of this
free energy to the metastable region gives rise to a complex free
energy, whose real part describes the equilibrium
properties of the metastable states and imaginary part their lifetime \cite{Langer67}. The free-energy functional or Hamiltonian~(\ref{H}), on the other
hand, is itself supposed to be a result of a constrained integration in (\ref{f}) of those fields over a spatial region of size $a$, or in terms of spatial Fourier
transform,
\begin{equation}
\phi({\bf k})=\int d{\bf x}\phi({\bf x})\exp(-i{\bf k\cdot
x}),\label{fourier}
\end{equation}
those fields whose wave numbers are larger than a momentum cutoff $\Lambda$, which is
proportional to $1/a$. This coarse-grained procedure leads to a functional that possesses
metastable and unstable states and is appropriate to describe their dynamical
properties \cite{Penrose,Langer74,Binder07}. Note that we have
used in Eq.~(\ref{fourier}) and shall use throughout the same symbol for both direct and Fourier transformed spaces.

\subsection{\label{phi3}Derived $\varphi^3$ model}
We shall study the FOPTs in model~(\ref{H}) and thus shall take $r<0$ throughout the paper. It is
well known then that there is a spatially uniform spontaneous
magnetization $M$, in the terminology of magnetism, below $T_c$ even
in the absence of $H$. Accordingly, it is essential to shift the
order parameter by $M$. In particular, let
\begin{equation}
\phi=M+\varphi,\label{pmvp}
\end{equation}
then,
\begin{eqnarray}
{\cal H} [\varphi] = V(rM^2/2 + gM^4/4! - HM) +\nonumber\\
\int d{\bf r} \left[
\frac{1}{2}\tau\varphi^2 +\frac{1}{3!}g_3\varphi^3+ \frac{1}{4!}g\varphi^4 + \frac{1}{2}(R\nabla \varphi)^2 -h\varphi\right],\label{ha}
\end{eqnarray}
where $V$ is the volume of the system,
\begin{equation}
\tau=r+\frac{1}{2}gM^2, \ \ \ h= H-rM-\frac{1}{3!}gM^3, \label{tauh}
\end{equation}
and
\begin{equation}
g_3=gM.\label{g3}
\end{equation}
We shall show later on that
the long-wavelength behavior of the system is dominated by the
leading $\varphi^3$ term in (\ref{ha}) \cite{zhongl05}. Therefore, neglecting
the $\varphi^4$ term, we have a derived $\varphi^3$ model
\cite{zhongl05}
\begin{equation}
{\cal H}_3 [\varphi] = \int d{\bf r} \left\{
\frac{1}{2}\tau\varphi^2 +\frac{1}{3!}g_3\varphi^3+
\frac{1}{2}[\nabla \varphi]^2 -h\varphi \right\}. \label{hp}
\end{equation}

Note that a $\varphi^3$ model has been noticed in an RG analysis of a mean-field spinodal fixed point \cite{Gunton78}, and has also been used in a nucleation theory
in systems with long-range interactions near the mean-field
spinodal point \cite{Klein83}. However, these theories are only of
mean-field nature. In contrast, we shall take fluctuations into
account.

\subsection{\label{dynamics}Dynamics}
Metastability is essentially kinetic in origin. In order to deal with it, we consider a phenomenological dynamics governed by the Langevin equation
\begin{equation}
\frac{\partial \phi }{\partial t} = - \lambda \frac{\delta {\cal H}[\phi ]}{\delta \phi } + \zeta ,
\label{de}
\end{equation}
i.e., Model A in the critical dynamics \cite{hohenberg}, with a Gaussian white noise $\zeta$ satisfying
\begin{eqnarray}
\langle\zeta({\bf x},t)\rangle  &=&  0, \nonumber\\
\langle\zeta({\bf x},t)\zeta({\bf x'},t')\rangle  &=&  2\lambda
\delta({\bf x}-{\bf x'})\delta(t-t'), \label{noise}
\end{eqnarray}
or equivalently, satisfying a
local functional probability distribution ${\cal D}\rho(\zeta)$
\begin{equation}
{\cal D}\rho(\zeta)={\cal D}\zeta\exp\left[-\frac{1}{4\lambda}\int
d{\bf x}dt\zeta^2({\bf x},t) \right], \tag{10$'$}\label{dxi}
\end{equation}
where $\lambda $ is a kinetic coefficient. The noise is supposed to mimic the effects on the order parameter of those integrated degrees of freedom, which have short relaxation times. Accordingly, there exists in principle also a coarse-grained time scale and hence a cutoff frequency. Nevertheless, we shall neglect this constraint in the following as we shall consider sufficiently long time universal behavior.

\section{\label{dft}Dynamical field theory}
The solution $\phi$ of the dynamic equations (\ref{de}) and
(\ref{noise}) depends on $\zeta$ and is thus stochastic and ought to
be averaged over the distribution (\ref{dxi}). The standard
practice is to formulate the problem in terms of a dynamical
field theory \cite{Janssen79,Janssen,Justin,Vasilev,Tauber,Folk}, because then
standard field-theoretic techniques become applicable. We shall briefly
repeat in this section the main steps leading to the dynamic action in Sec.~\ref{da} and collect the definitions of the generating functionals for connected Green functions and vertex functions as well as their relationship in Sec.~\ref{cgfvf}. The formulation of the theory in a supersymmetry form is left to Appendix~\ref{sa}, which serves also to indicate the supersymmetry origin of some of the relationship.

\subsection{\label{da}Dynamic action}
A convenient method to preform the averages is to calculate the
generating functional defined as
\begin{equation}
Z[J]=\left\langle\exp\left[\int d{\bf x}dtJ({\bf x},t)\phi({\bf
x},t)\right]\right\rangle, \label{Z}
\end{equation}
where the angle brackets denote average over $\zeta$. Then
quantities of interest such as the magnetization and $n$-point
correlation functions are readily obtainable by derivatives with
respect to the external source $J$,
\begin{eqnarray}
C_1(x)&\equiv&\langle\phi(x)\rangle =
\left.\frac{1}{Z[0]}\frac{\delta Z}{\delta
J(x)}\right|_{J=0},\label{m}\\
C_n(x_1,\dots,x_n)&\equiv&\langle\phi(x_1)\dots\phi(x_n)\rangle\nonumber\\
&=&\left. \frac{1}{Z[0]}\frac{\delta^n Z}{\delta J(x_1)\dots\delta
J(x_n)}\right|_{J=0}, \label{cf}
\end{eqnarray}
respectively, where we have used $x_n$ to denote $({\bf
x}_n,t_n)$.

The condition that $\phi$ must be the solution of Eq.~(\ref{de})
can be fulfilled by inserting in Eq.~(\ref{Z}) a Dirac delta
function, which may in turn be represented by a functional Fourier
transform by introducing an auxiliary response field $\tilde {\phi
}$ \cite{martin}. So,
\begin{eqnarray}
Z[J]=\int {\cal D}\rho(\zeta){\cal D}\phi {\cal D}\tilde{\phi}{\cal
J}\nonumber\\
\exp\left\{\int d{\bf x}dt\left[J\phi + \tilde{\phi}\left( \zeta-
\frac{\partial \phi }{\partial t} - \lambda \frac{\delta {\cal
H}}{\delta \phi }\right)\right]\right\},
\end{eqnarray}
where the integration over $\tilde{\phi}$ runs along the imaginary
axis, and ${\cal J}$, given formally by
\begin{equation}
{\cal J}= \det\left[\frac{{\cal D}\zeta}{{\cal D}\phi}\right]=\det\left[\left( \frac{\partial}{\partial t} + \lambda
\frac{\delta^2 {\cal H}}{\delta \phi^2 }\right)\right],
\label{jacobi}
\end{equation}
is the functional Jacobian for transforming
to the integration over $\phi$ with $\det$ denoting the determinant. Integrating over the noise leads to
\begin{equation}
Z[J,\tilde{J}]=\int {\cal D}\phi {\cal D}\tilde{\phi}{\cal
J}\exp\left[-{\cal L}+\int d{\bf x}dt (J\phi
+\tilde{J}\tilde{\phi})\right] \label{z'}
\end{equation}
with the action ${\cal L}$ given by
\begin{equation}
{\cal L}=\int d{\bf x}dt\left[\tilde{\phi}\left( \frac{\partial \phi }{\partial t} +
\lambda \frac{\delta {\cal H}}{\delta \phi }\right)-\lambda
\tilde{\phi}^2\right] . \label{Lj}
\end{equation}
To Eq.~(\ref{z'}) we have inserted another source $\tilde{J}$
conjugate to $\tilde{\phi}$ such that response functions which are average of
$\tilde{\phi}$ and $\phi$ fields can be calculated by the method similar to
Eq.~(\ref{cf}). In particular, the two-point response function
$G_{11}(x,x')=\langle\phi(x)\tilde{\phi}(x')\rangle$ is
\begin{equation}
G_{11}(x,x')=\left.\frac{1}{Z[0,0]}\frac{\delta^2
Z[J,\tilde{J}]}{\delta J(x)\delta
\tilde{J}(x')}\right|_{\!\scriptsize{\begin{array}{l}J=0\\\tilde{J}=0\end{array}}}
=\frac{1}{\lambda}\frac{\delta\langle\phi(x)\rangle}{\delta
H(x')} \label{rf2}
\end{equation}
by noting that an external field is equivalent to the source
$\tilde{J}$ up to $\lambda$. The last equality of Eq.~(\ref{rf2}) expresses the
response to an external field and thus is the origin for calling $\tilde{\phi}$
the response field. The response field in
Eq.~(\ref{Lj}) may be integrated out by a Gaussian integral, but
the resultant functional is nonlinear and inconvenient for
treatment.

At this point, there are two different methods to continue in
tackling the Jacobian (\ref{jacobi}), giving rise therefore
to two different forms of action. We shall proceed with the usual one and leave the more exotic one to Appendix~\ref{sa}.

The usual method is to note that the determinant \cite{Janssen79,Justin,Tauber}
\begin{equation}
{\cal J} = \exp \left[\Theta(0)\lambda \int d{\bf
x}dt\frac{\delta^2 {\cal H}}{\delta
\phi^2}\right]\equiv\exp\left(-{\cal L}'\right),
\end{equation}
where the Heaviside step function defined as
\begin{equation}
\Theta(t)=\left\{\begin{array}{ll}1,& t>0\\-1,&t<0\end{array}\right.
\end{equation}
has to be assigned a special value $\Theta(0)=1/2$ for
consistency in the continuum limit of a symmetrized discretization
of the Langevin equation \cite{Janssen,Justin,Tauber}. One therefore has a
dynamical field theory with a total action ${\cal L}_{\rm tot}={\cal
L}+{\cal L}'$. Standard field theoretical methods can then be
utilized to compute relevant response functions. In
particular, Feynman rules for perturbation expansions may be
defined (see Sec.~\ref{pert} below). One can then prove that diagrams with a
closed response loop given by a contraction of $\phi$ and
$\tilde{\phi}$ at the same spatial-temporal point (i.e., both $\phi$ and
$\tilde{\phi}$ come from the same term
$\lambda\tilde{\phi}\delta{\cal H}/\delta\phi$ in Eq.~(\ref{Lj}))
give just $\lambda G(0)\delta^2{\cal H}/\delta\phi^2=-{\cal L}'$ and thus just
cancel the Jacobian exactly in each order of the perturbation
expansions because $G(0)=\Theta(0)$ (see Eq.~(\ref{gkt}) below). As a consequence,
one may simply ignore the Jacobian by choosing $\Theta(0)=0$ \cite{Janssen79,Janssen,Tauber},
which corresponds to a forward discretization, and meanwhile,
excludes closed response loops from the perturbation expansions.
Note that in this way, in the ensuing dynamical field theory,
\begin{equation}
Z[J,\tilde{J}]=\int {\cal D}\phi {\cal
D}\tilde{\phi}\exp\left[-{\cal L}+\int d{\bf x}dt (J\phi
+\tilde{J}\tilde{\phi})\right] \label{zjj}
\end{equation}
with the dynamic action given by Eq.~(\ref{Lj}), causality becomes
automatically implemented because \cite{Bausch,Janssen79}
\begin{equation}
\langle\phi(t_1)\dots\phi(t_n)\tilde{\phi}(t'_1)\dots\tilde{\phi}(t'_{n'})\rangle=0,\
\ {\rm if\ one\ }t'_{i'}>{\rm all}\ t_i.\label{causality}
\end{equation}
In particular,
\begin{equation}
\langle\tilde{\phi}(t'_1)\dots\tilde{\phi}(t'_{n'})\rangle=0.\label{g0n}
\end{equation}

\subsection{\label{cgfvf}Connected Green functions and vertex functions}
From $Z[J,\tilde{J}]$, Eq.~(\ref{zjj}), or, the partition function
in statistical mechanics as it may be regarded as defining the
statistical weight of a configure $(\phi, \tilde{\phi})$, one
defines another generating functional $W[J,\tilde{J}]$, the minus free
energy, by
\begin{equation}
W[J,\tilde{J}]=\ln Z[J,\tilde{J}] \label{w}
\end{equation}
for connected correlation and response functions, or the cumulants
\begin{eqnarray}
&&G_{nn'}^c(x_1,\dots,x_n,x'_1,\dots,x'_{n'})\nonumber\\
&\equiv&\langle\phi(x_1)\dots\phi(x_n)\tilde{\phi}(x'_1,\dots,x'_{n'})\rangle_c\nonumber\\
&=&\left.\frac{\delta^{n+n'} W[J,\tilde{J}]}{\delta
J(x_1)\dots\delta
J(x_n)\delta\tilde{J}(x'_1)\dots\tilde{J}(x'_{n'})}\right|_{\!\scriptsize{\begin{array}{l}J=0\\\tilde{J}=0\end{array}}},
\label{gcd}
\end{eqnarray}
which are generally deviations and their moments. For example, the
connected two-point correlation function
\begin{eqnarray}
G_{20}^c(x,x')&=&\langle\left[\phi(x)-\langle\phi\rangle\right]\left[\phi(x')-\langle\phi\rangle\right]\rangle\nonumber\\
&=&C_2(x,x')-\langle\phi\rangle^2\equiv C(x,x')\label{c2}
\end{eqnarray}
is the second moment of the deviation $\phi-\langle\phi\rangle$. Accordingly, one may construct the correlation and response
functions from their connected counterparts, whose Feynman
diagrams are fewer in number in high orders in perturbation expansions.

The connected diagrams contain one-line reducible and irreducible
diagrams that depend on whether they are connected or not when
an internal line of the diagrams is cut. The former is products of
the latter, which can be proved to be generated by the Gibbs
free energy $\Gamma$ through a Legendre transformation
\begin{equation}
\Gamma[\langle\tilde{\phi}\rangle,\langle\phi\rangle]=-W[J,\tilde{J}]+\int d{\bf x}dt\left(
J\langle\phi\rangle +\tilde{J}\langle\tilde{\phi}\rangle\right),\label{ga}
\end{equation}
where $\langle\phi\rangle$ and $\langle\tilde{\phi}\rangle$ are averages in the presence of $J$ and $\tilde{J}$ given by
\begin{equation}
\langle\phi\rangle\equiv\frac{\delta W}{\delta J},\ \ \
\langle\tilde{\phi}\rangle\equiv\frac{\delta W}{\delta \tilde{J}}.\label{pw}
\end{equation}
Then, from Eqs.~(\ref{ga}) and (\ref{pw}),
\begin{equation}
J=\frac{\delta \Gamma}{\delta \langle\phi\rangle},\ \ \ \tilde{J}=\frac{\delta
\Gamma}{\delta \langle\tilde{\phi}\rangle},\label{jg}
\end{equation}
and the one-line irreducible vertex function $\Gamma_{n'n}$
is
\begin{eqnarray}
\Gamma_{n'n}(x'_1,\dots,x'_{n'},x_1,\dots,x_n)=\quad\nonumber\\
\left.\frac{\delta^{n'+n}
\Gamma[\langle\tilde{\phi}\rangle,\langle\phi\rangle]}{\delta\langle\tilde{\phi}(x'_1)\rangle \dots\delta\langle\tilde{\phi}(x'_{n'})\rangle \delta\langle\phi(x_1)\rangle\dots\delta
\langle\phi(x_n)\rangle}
\right|_{\!\scriptsize{\begin{array}{l}J=0\\\tilde{J}=0\end{array}}}\!.\quad
\label{g}
\end{eqnarray}
In particular, differentiating Eq.~(\ref{jg}) with $J$ and $\tilde{J}$ and using Eqs.~(\ref{gcd}) and (\ref{c2}) give
\begin{eqnarray}
\Gamma_{02}(x_1,x_2)=0,\label{gamma02}\qquad\\
\int dx\Gamma_{11}(x_1,x)G_{11}(x,x_2)=\delta(x_1-x_2),\label{g11}\qquad\\
\int dx\Gamma_{20}(x_1,x)G_{11}(x,x_2)=- \int dxG_{20}^c(x_1,x)\Gamma_{11}(x,x_2)\nonumber\\
=- \int dx C(x_1,x)\Gamma_{11}(x,x_2),\ \label{g20}\qquad
\end{eqnarray}
where the first equation is a result of Eq.~(\ref{g0n}) and use has been made of $G_{11}^c=G_{11}$ due to Eq.~(\ref{g0n}). In addition, the fluctuation-dissipation theorem
\begin{equation}
\lambda [G_{11}(x,x')-G_{11}(x',x)]=-\frac{\partial C(x,x')}{\partial t}\label{fdt}
\end{equation}
holds relating the response function to the correlation function \cite{deker,Justin,de,Janssen79,Vasilev,Folk,Tauber}. Moreover, dynamic response functions converge to their corresponding static correlation functions at long times \cite{de,Janssen79,Justin,Vasilev}. In fact, these are consequences of the supersymmetry in the theory \cite{Justin}.

\section{\label{mft}Mean-field theory}

In this section, we study the mean-field theory of a field-driven
FOPT and show that it is controlled by an instability point \cite{zhongl05}, which is the spinodal point in the mean-field theory.

\subsection{$\varphi^3$ model and instability point}
The mean-field theory is the lowest order of a saddle-point
approximation to the free energy (\ref{f}). Assuming that the
saddle point lies at a uniform order parameter $M$, we have the free energy density
\begin{equation}
F={\cal F}/V={\cal H}/V=\frac{1}{2}rM^2 + \frac{1}{4!}gM^4 - HM.\label{mf4}
\end{equation}
The dynamics then reduces to
\begin{equation}
\frac{dM}{dt} = - \lambda \left(rM + \frac{1}{3!}gM^3  - H
\right), \label{mf}
\end{equation}
from Eqs.~(\ref{de}) and (\ref{mf4}).
It is well known that there is a mean field critical point at $r = 0$ and
$H = 0$ with its mean field critical exponents collected in Table~\ref{mfexp} for later comparison. For $r < 0 $, on the other hand, there is an equilibrium FOPT at $H = 0$ between the two phases with $M =\pm
\sqrt { - 6r / g} \equiv \pm M_{e}$ for each given $r$ and $g$.
As there is no fluctuation, the FOPT cannot take place at the equilibrium transition point at $H=0$ because there is a free-energy barrier of $|F(0)-F(M_e)|=3r^2/2g$ between the two phases. Rather, it can only take place beyond the point where the barrier vanishes. This is in fact the spinodal point at which both the first and the second derivatives of $F$ with respect to $M$ vanish.
\begin{table}
\caption{\label{mfexp} Mean-field critical and instability exponents.}
\begin{ruledtabular}
\begin{tabular}{ccccccccccc}
Theory&$d_c$&$\beta$&$\delta$&$\gamma$&$\alpha$&$\nu$&$\eta$&$z$&$n_H$&$n_m$\\
\hline
$\phi^4$ &4&$\frac{1}{2}$&3&1 &0   &$\frac{1}{2}$&0&2&$\frac{3}{5}$&$\frac{1}{5}$\\
$\varphi^3$ &6&1            &2&1 &$-1$&$\frac{1}{2}$&0&2&$\frac{2}{3}$&$\frac{1}{3}$\\
\end{tabular}
\end{ruledtabular}
\end{table}

To see the essence of this transition, let us set $M=M_s+m(t)$ in
Eq.~(\ref{mf}), which becomes
\begin{equation}
\frac{dm}{dt} = - \lambda \left( {\tau m + \frac{1}{2}gM_s m^2
+\frac{1}{3!} gm^3 - h } \right), \label{sm}
\end{equation}
where $M_s$ is a constant, and
\begin{equation}
\tau = r+\frac{1}{2}gM_s^2, \qquad h = H - rM_s-\frac{1}{3!}gM_s^{3},\label{sp}
\end{equation}
which is just Eq.~(\ref{tauh}) at $M_s$. One finds a pair of the spinodal points which lie at
\begin{equation}
\tau_s= 0, \qquad h_s= 0, \label{mfs}
\end{equation}
or from Eq.~(\ref{sp}), at $M_{s}= \pm \sqrt { - 2r /g} $ and $H_{s} = \mp (2r / 3)\sqrt{ - 2r / g} $. Accordingly, Eq.~(\ref{sp}) can be expressed as
\begin{equation}
\tau = c_1(T-T_s), \qquad h = H - H_s \label{sps}
\end{equation}
with $T_s=T_c-gM_s^2/2c_1$ at either point. Comparing with Eq.~(\ref{mf}), one sees that each point at its associated transition plays a similar role to the critical point. The only difference is that one has a quadratic term in Eq.~(\ref{sm}). However, in the vicinity of the spinodal point, $m$ is  small. This term overwhelms the cubic one and thus controls the dynamics.
Therefore, upon neglecting the cubic term in
Eq.~(\ref{sm}), the dynamics in the vicinity of the spinodal
point becomes
\begin{equation}
\frac{dm}{dt} = - \lambda \left( {\tau m + \frac{1}{2}gM_s m^2 - h } \right), \label{m3}
\end{equation}
which is governed by the derived $\varphi^3$ model, Eq.~(\ref{hp}), whose corresponding mean-field free energy density is
\begin{equation}
F_{3} =  \frac{1}{2}\tau m ^2 + \frac{1}{3!}gM_sm ^3 - hm,
\label{fcube}
\end{equation}
and each spinodal point, Eq.~(\ref{mfs}), now becomes the
instability point of the state $m=0$, because at these points,
the system becomes unstable as can be seen in Fig.~\ref{mff}(a).

Note that if we neglect the quadratic term in Eq.~(\ref{sm}), we return to the $\phi^4$ theory. Only the critical point is now replaced by one of the two spinodal points. This is the theory of pseudo-critical phenomena in which usual critical behavior emerges at
the spinodal point instead of the critical point \cite{Saito}. However, in this case, the
critical fixed point is unstable against the neglected quadratic
term and only crossover may be observed \cite{Saito}.

\subsection{Mean-field static instability exponents}
As the free energy of the $\varphi^3$ model is known, one can then derive its associated exponents similar to the $\phi^4$ mean-field model. We call them instability exponents or even spinodal exponents as they are associated with the instability point or spinodal point. One will see that they are exactly the counterparts of the critical exponents. Accordingly, we shall use identical symbols with the critical ones.

At equilibrium, Eq.~(\ref{m3}) leads to the equation of state
\begin{equation}
\tau m + \frac{1}{2}gM_s m^2 = h. \label{eos}
\end{equation}
As a result,
\begin{equation}
m=\left\{\begin{array}{ll}0,& \tau>0,\\-2\tau/gM_s\sim(T_s-T)^{\beta},& \tau<0
\end{array}
\right.\label{m3eq}
\end{equation}
for $h=0$. So, $m$ changes continuously to zero with $\beta=1$. Also, as $h\sim m^2$ at $\tau=0$, $\delta=2$. Differentiating Eq.~(\ref{eos}) gives rise to
\begin{equation}
\chi=\left.\frac{\partial m}{\partial h}\right|_{h=0}=\left\{\begin{array}{ll}\tau^{-1}\sim(T-T_s)^{-\gamma},& \tau>0,\\(-\tau)^{-1}\sim(T_s-T)^{-\gamma'},& \tau<0
\end{array}
\right.
\end{equation}
for the susceptibility. So, $\gamma=\gamma'=1$, which are identical to their critical counterparts. However, the amplitude ratio between the susceptibility above and below $\tau=0$ is now 1 instead of $2$. The free energy density for the two solutions is
\begin{equation}
F_3=\left\{\begin{array}{ll}0,& \tau>0,\\2\tau^3/3g^2M_s^2, & \tau<0.
\end{array}
\right.
\end{equation}
Thus the specific heat is
\begin{equation}
C=-T\frac{\partial^2F_3}{\partial T^2}=\left\{\begin{array}{ll}0,& \tau>0,\\-4\tau c_1^2T/g^2M_s^{2}\sim(T_s-T)^{-\alpha},& \tau<0,
\end{array}
\right.\label{sh}
\end{equation}
which again changes continuously from one state to the other with $\alpha=-1$. These instability exponents are also collected in Table~\ref{mfexp}.

\begin{figure}[t]
\centerline{\epsfig{file= 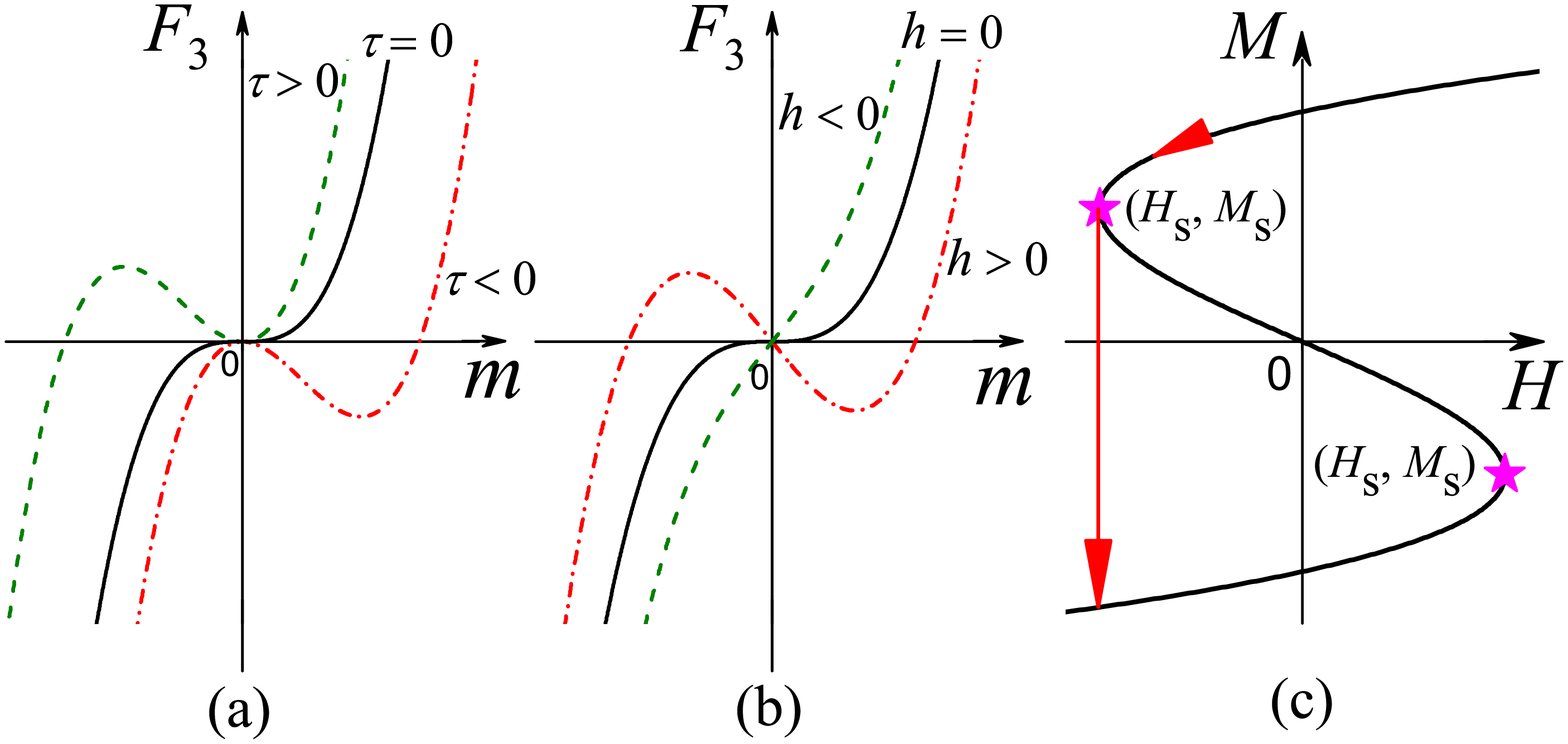,height=6cm,width=9cm}}
\vskip -2 cm\caption{\label{mff}(Color online) $\varphi^3$ mean-field free-energy density for (a) different $\tau$s at $h=0$ and (b) different $h$s at $\tau=0$. (c) Generic hysteresis of the $\phi^4$ model in an external field $H$. The arrows depict the transition that can be described by (b). The stars mark the spinodal/instability points.}
\end{figure}
A caveat is needed here. One sees from Eq.~(\ref{m3eq}) and Fig.~\ref{mff}(a) that the instability exponents just derived describe the continuous transition from the state with $m=-2\tau/gM_s$ to that with $m=0$. As the specific heat is linear in $\tau$ from Eq~(\ref{sh}), this is in fact a third-order phase transition in the classical classification \cite{Ehrenfest}. Does this transition have anything to do with the FOPT we attempt to study? Our answer is yes. This can be seen from Fig.~\ref{mff}(b). As mentioned, there are two transitions associated with two spinodal points (marked by starts in Fig.~\ref{mff}(c)) which are related by inversion symmetry. Without loss of generality, let us consider the one with positive $M_s$ and negative $H_s$. The FOPT is then driven by increasing $H$ in the direction opposite to $M_s$ as shown in Fig.~\ref{mff}(c). This corresponds to a positive $gM_s$ and $h$ changes from positive to negative, as Fig.~\ref{mff}(b) shows. As long as $0>H>H_s$ and hence $h=H-H_s>0$, the system keeps in the metastable state with $M>0$. This corresponds to the situation in which $m$ resides in the well. As $h$ decreases, the well becomes shallower and closer to the origin. Exactly at $h=0$, the system lies at the transition point and becomes unstable. When $h<0$, the positive $m$ state loses its stability and would fall to negative infinity eventually. In practice, of course, the system restabilizes at a negative $M$ due to the neglected quartic term in the free energy when $m$ gets large (negative). Thus, the exponents are in fact the properties of the transition point at $h=0$ rather than the well at $h>0$. This becomes apparent in the RG theory of the critical phenomena and also of FOPTs as will be seen later on. In this theory, the exponents are properties of a fixed point. Only the critical point of a real system converges upon renormalization to the fixed point, not the equilibrium phase. This indicates that the RG is insensitive to the minima which are a finite distance away \cite{Amit76}, although this argument which arose from the RG study of the Potts model whose Landau mean-field theory predicts an FOPT instead of a continuous transition \cite{Harris75} was argued to be doubtable \cite{Amit76}. Nevertheless, we shall show in this paper that the instability point does describe the FOPT. However, in the FOPT we considered, the system falls into the unstable left side instead of the right well. This may be the reason why the fixed point we shall find below is imaginary \cite{zhongl05}.

\subsection{\label{mfhys}Mean-field hysteresis exponents and their verification}
We now show that the mean-field instability exponents are indeed relevant to the FOPTs considered.

\begin{figure}[t]
\centerline{\epsfig{file= 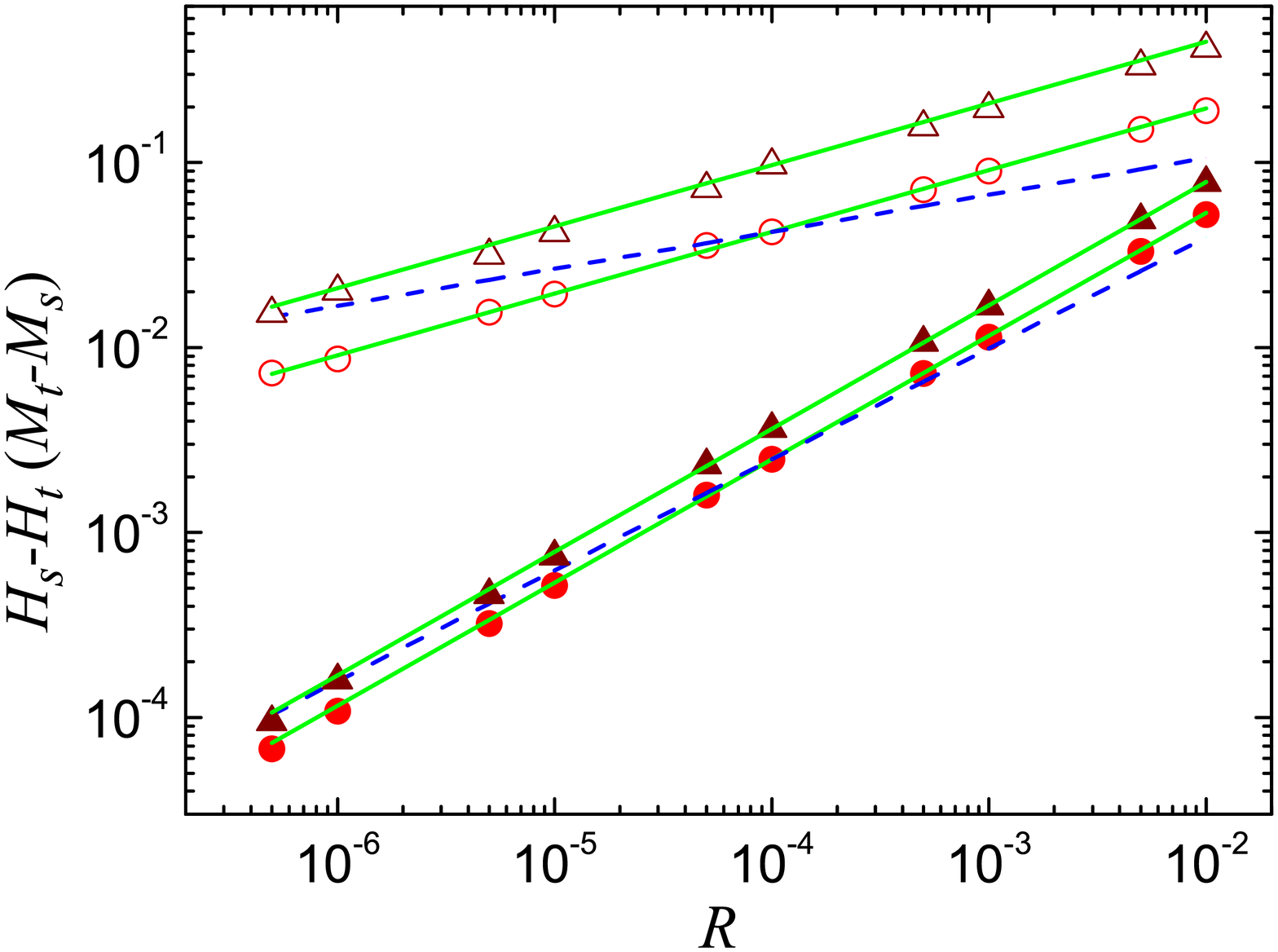,height=6cm,width=9cm}}
\caption{\label{mfn}(Color online) Mean-field transition fields $H_s-H_t$ (filled symbols) and moments $M_t-M_s$ (opened symbols) versus $R$. Full lines through filled symbols and opened symbols are lines of slopes $2/3$ and $1/3$, respectively, through the data points at $R=10^{-4}$. The two dashed lines have slopes $3/5$ (lower) and $1/5$ (upper) for comparison. Note that the symbols have small vertical variations due to the interpolations in extracting the points but can never fit the dashed lines. The parameters chosen are $r=-1$, $\lambda=1$, and $g=1$ (circles) and $g=0.1$ (triangles).}
\end{figure}

In order to probe these exponents, we shall utilize the method of finite-time scaling proposed in critical phenomena \cite{gong,zhongintech}, a method in which $H$ is swept linearly through the instability point. To be specific, we consider again the transition depicted in Fig.~\ref{mff}(c) and assume that
\begin{equation}
h=-Rt, \quad {\rm or,}\quad H=H_s-Rt, \label{hRt}
\end{equation}
where $R$ is a constant. This form of driving implies choosing $t=0$ at $h=0$. In fact, one can start sweeping the field from anywhere sufficiently far away from $H_s$. For example, $H=H_0-Rt$. However, $h=0$ always shifts the time origin to the time at $H_s$, namely, $h=-R(t-t_s)$, which recovers Eq.~(\ref{hRt}) upon setting $t_s=0$. The linear driving imposes on the system an effective finite time scale which is proportional to $R^{-1}$. When it is shorter than the correlation time, finite-time scaling follows in close analogy to the occurrence of finite-size scaling when the size of the system is smaller than its correlation length. We shall show below in Sec.~\ref{fts} that at the instability point $\tau=0$ and $h=0$, $m$ follows a finite-time scaling form
\begin{equation}
m(h,R)=R^{\beta/\nu r_H}f(-hR^{-\beta\delta/\nu r_H}), \label{mscaling}
\end{equation}
where
\begin{equation}
r_H=z+\beta\delta/\nu \label{sl}
\end{equation}
is the RG eigenvalue associated with $R$, $z$ and $\nu$ are instability exponents, and $f$ is a scaling function. Using the mean-field instability exponents derived in the last section and those that will be derived in next section, all collected in Table~\ref{mfexp}, one finds that Eq.~(\ref{mscaling}) becomes
\begin{equation}
m(h,R)=R^{1/3}f(-hR^{-2/3}). \label{mscalingn}
\end{equation}
This means that at $m=0$ or $M=M_s$, the transition field
\begin{equation}
h_t\equiv H_t-H_s=c_2R^{2/3}\sim R^{n_H} \label{mht}
\end{equation}
and at $h=0$, the transition moment
\begin{equation}
m_t(0,R)\equiv M_t-M_s=f(0)R^{1/3}\sim R^{n_m}, \label{mmt}
\end{equation}
where $c_2$ is the root of $f$, i.e., $f(-c_2)=0$ and $n_H=\beta\delta/\nu r_H=2/3$ and $n_m=\beta/\nu r_H=1/3$ are hysteresis exponents \cite{zhongl05}. All these results are born out by direct numerical solutions of Eq.~(\ref{mf}) shown in Fig.~\ref{mfn}. Note that if, instead of the $\varphi^3$ theory, the original $\phi^4$ theory governed the transition, the critical exponents listed in Table~\ref{mfexp} would give $n_H=3/5$ and $n_m=1/5$ \cite{zhong06}, which can be ruled out from Fig.~\ref{mfn}. This indicates that the theory of pseudo-critical phenomena \cite{Saito} cannot describe at least the mean-field theory.

In fact, the above results can even be proved analytically. Eq.~(\ref{m3}) is a kind of the Ricatti equations that can be solved analytically \cite{jung}. For the driving~(\ref{hRt}),
\begin{equation}
m=\frac{2}{g\lambda y}\frac{dy}{dt}, \qquad s=\sqrt[3]{g\lambda^2R/2}t \label{mst}
\end{equation}
transform Eq.~(\ref{m3}) into the Airy equation
\begin{equation}
\frac{d^2y}{ds^2}=sy
\end{equation}
which is solved by the Airy functions $Ai$ and $Bi$ as
\begin{equation}
y=c_3Ai(s)+c_4Bi(s),\label{airy}
\end{equation}
where $c_3$ and $c_4$ are constants to be determined by initial conditions. Consequently,
\begin{equation}
m(t)=\sqrt[3]{\frac{4R}{g^2\lambda}}\frac{c_3Ai'(\sqrt[3]{g\lambda^2R/2}t)+c_4Bi'(\sqrt[3]{g\lambda^2R/2}t)} {c_3Ai(\sqrt[3]{g\lambda^2R/2}t)+c_4Bi(\sqrt[3]{g\lambda^2R/2}t)}\label{mas}
\end{equation}
from Eqs.~(\ref{mst}) and (\ref{airy}), where the primes indicate a derivative with respect to the argument. Therefore, one recovers Eq.~(\ref{mht}) from Eq.~(\ref{mst}) with $c_2=s_t/\sqrt[3]{g\lambda^2/2}$ at $s=s_t$ at which $m=0$ and Eq.~(\ref{mmt}) from Eq.~(\ref{mas}) at $h=Rt=0$ with $f(0)=\sqrt[3]{4/g^2\lambda}[c_3Ai'(0)+c_4Bi'(0)]/[c_3Ai(0)+c_4Bi(0)]$.

These mean-field results therefore shows clearly that the dynamic scaling of
hysteresis in the FOPT of the $\phi^4$ model is determined by the instability point of the derived $\varphi^3$ model.

\section{\label{gauss}Gaussian theory}
In this section, we consider the Gaussian theory for the $\varphi^3$ model. This theory, which can be solved analytically, also describes fluctuations in the state $m=0$. As a result, the other instability exponents $\nu$, $z$, and $\eta$ can all be derived. More importantly, we shall see clearly that at the instability point, the correlation length and the correlation time both diverge similar to their critical counterparts.

To this end, we use the dynamic action, Eq.~(\ref{Lj}), which, for ${\cal H}_3$ in the absence of the cubic term and $h$, is
\begin{widetext}
\begin{equation}
{\cal L}_0=\int d{\bf x}dt\left\{\tilde{\varphi}\left[ \frac{\partial \varphi }{\partial t} +
\lambda \left(\tau\varphi-\nabla^2\varphi\right) \right]-\lambda
\tilde{\varphi}^2\right\}
=\int d\overline{{\bf k}}d\overline{\omega} \left\{[i\omega+\lambda(\tau+{\bf k}^2)]\tilde{\varphi}({\bf k},\omega)\varphi(-{\bf k},-\omega)
-\lambda |\tilde{\varphi}({\bf k},\omega)|^2\right\}
, \label{L3}
\end{equation}
where we have written $\tilde{\varphi}$ in place of $\tilde{\phi}$ for uniformity and the inverse Fourier transform is defined as
\begin{equation}
\varphi({\bf x},t)=\int d\overline{{\bf k}}d\overline{\omega}\varphi({\bf
k},\omega)\exp(i{\bf k\cdot x}-i\omega t)\label{ifou}
\end{equation}
with
\begin{equation}
\int d\overline{{\bf k}}d\overline{\omega}\equiv\frac{1}{(2\pi)^{d+1}}\int d{\bf k}d\omega,
\end{equation}
$d$ being the space dimensionality. Note that, as we used a finite volume, we should in principle replace $[1/(2\pi)^{d}]\int d{\bf k}$ by a discrete sum $(1/V)\sum_{\bf k}$ for consistency. However, only in the thermodynamic limit in which $V\rightarrow\infty$ and hence the sum becomes the continuous integral can singularity emerges. So, we still use the integral representation. Integrating out both $\tilde{\varphi}$ and $\varphi$, one finds
\begin{equation}
W[J,\tilde{J}]\sim\frac{1}{2}\int d\overline{{\bf k}}d\overline{\omega}\left[G({\bf k},-\omega)\tilde{J}(-{\bf k},-\omega)J({\bf k},\omega)+G({\bf k},\omega)\tilde{J}({\bf k},\omega)J(-{\bf k},-\omega)+C({\bf k},\omega)J(-{\bf k},-\omega)J({\bf k},\omega)\right] \label{wfr}
\end{equation}
from Eq.~(\ref{w}), where we have neglected some irrelevant constants,
\begin{equation}
G({\bf k},\omega)=\frac{1}{-i\omega+\lambda(\tau+{\bf k}^2)},
\label{gc}
\end{equation}
is related to the two-point response function defined in Eq.~(\ref{gcd}) by
\begin{equation}
G_{11}({\bf k}_1,\omega_1;{\bf k}_2,\omega_2)=\int dx_1dx_2G_{11}(x_1,x_2)e^{i\sum_{j=1}^{2}(\omega_j t_j-{\bf k}_j\cdot{\bf x}_j)}
=(2\pi)^{d+1}\delta({\bf k}_1+{\bf k}_2)\delta(\omega_1+\omega_2)G({\bf k}_1,\omega_1),\label{gk2k}
\end{equation}
the last equality being resulted from translational invariance, and
\begin{equation}
C({\bf k},\omega)=2\lambda G({\bf k},\omega)G({\bf k},-\omega)=\frac{2\lambda }{\omega^2+\lambda^2(\tau+{\bf k}^2)^2},\label{c2g}
\end{equation}
is related to the two-point correlation function defined in Eq.~(\ref{c2}) through a relation similar to Eq.~(\ref{gk2k}). Equations.~(\ref{gc}) and (\ref{c2g}) satisfy of course the fluctuation-dissipation theorem, Eq.~(\ref{fdt}),
\begin{equation}
\lambda [G({\bf k},\omega)-G({\bf k},-\omega)]=i\omega C({\bf k},\omega)\label{fdtf}
\end{equation}
in its Fourier transformed form. From Eq.~(\ref{wfr}), one also has
\begin{equation}
G_{02}^c({\bf k},\omega)=0,
\end{equation}
which is consistent with Eq.~(\ref{g0n}). As a result, Eqs.~(\ref{g11}) and (\ref{g20}) become
\begin{equation}
\Gamma_{11}({\bf k},\omega)=G(-{\bf k},-\omega)^{-1}=i\omega+\lambda(\tau+{\bf k}^2),\qquad
\Gamma_{20}({\bf k},\omega)=-\frac{C({\bf k},\omega)}{G({\bf k},\omega)G(-{\bf k},-\omega)}=-2\lambda.\label{g11g}
\end{equation}
In fact, since
\begin{equation}
\langle\varphi({\bf k},\omega)\rangle=G({\bf k},\omega)\tilde{J}({\bf k},\omega)+C({\bf k},\omega)J({\bf k},\omega),\qquad
\langle\tilde{\varphi}({\bf k},\omega)\rangle=G({\bf k},-\omega)J({\bf k}\omega)
\end{equation}
from Eq.~(\ref{pw}), solving out $J$ and $\tilde{J}$ and using Eq.~(\ref{ga}), one finds
\begin{equation}
\Gamma=\frac{1}{2}\int d\overline{{\bf k}}d\overline{\omega}\left[G({\bf k},\omega)^{-1}\langle\tilde{\varphi}(-{\bf k},-\omega)\rangle\langle\varphi({\bf k},\omega)\rangle+G({\bf k},-\omega)^{-1}\langle\tilde{\varphi}({\bf k},\omega)\rangle\langle\varphi(-{\bf k},-\omega)\rangle-2\lambda \langle\tilde{\varphi}(-{\bf k},-\omega)\rangle\langle\tilde{\varphi}({\bf k},\omega)\rangle\right],
\end{equation}
\end{widetext}
which recovers Eq.~(\ref{g11g}) correctly by Eq.~(\ref{g}). Combining Eqs.~(\ref{fdtf}) and (\ref{g11g}), one can write the fluctuation-dissipation theorem in the form
\begin{equation}
i\omega\Gamma_{20}({\bf k},\omega)=\lambda[\Gamma_{11}({\bf k},-\omega)-\Gamma_{11}({\bf k},\omega)],\label{fdt20}
\end{equation}
or,
\begin{equation}
\Gamma_{20}({\bf k},\omega)=-\frac{2\lambda}{\omega}{\rm Im}\Gamma_{11}({\bf k},\omega),\label{fdt20i}
\end{equation}
which is of course satisfied by the Gaussian results.

Equation~(\ref{gc}) implies a relaxation or correlation time $t_{\rm eq}({\bf k})$ satisfying
\begin{equation}
\lambda t_{\rm eq}({\bf k})=(\tau+{\bf k}^2)^{-1}=\xi^zf_1({\bf k}\xi), \label{teq}
\end{equation}
where
\begin{equation}
\xi=\tau^{-1/2}\sim\tau^{-\nu} \label{xi}
\end{equation}
is a correlation length and
\begin{equation}
f_1(x)=\frac{1}{1+x^2}
\end{equation}
is a scaling function with the instability exponents $\nu=1/2$ and $z=2$. Indeed, the temporal Fourier transform of Eq.~(\ref{gc}) gives
\begin{equation}
G({\bf k},t)=\Theta(t)e^{-\lambda(\tau+{\bf k}^2)t}=\Theta(t)e^{-t/t_{\rm eq}({\bf k})} \label{gkt}
\end{equation}
from the residue theorem by noting that $G({\bf k},\omega)$ has a pole in the negative imaginary $\omega$ axis. So, the correlation dies out after the time $t_{\rm eq}({\bf k})$. Consequently, $G({\bf k},\omega)$ can be cast in a scaling form \cite{hohenberg}
\begin{equation}
G({\bf k},\omega)=\xi^{2-\eta}f_2({\bf k}\xi,(\omega/\lambda)\xi^z)/\lambda \label{gcs}
\end{equation}
with $\eta=0$ and the scaling function
\begin{equation}
f_2(x,y)=\frac{1}{1-iy+x^2},
\end{equation}
where the extra $\lambda$ arises from the time unit. So can $C({\bf k},\omega)$ and $\Gamma_{11}({\bf k},\omega)$. Note that $t_{\rm eq}({\bf k})$ diverges for $|{\bf k}|\rightarrow 0$ and $\tau=0$ similar to $\xi$.

As pointed out above, there exist relations between dynamic and static functions. One sees from Eqs.~(\ref{gc}) and (\ref{g11}) that at $\omega=0$ or the long-time limit, the dynamic functions equal the static ones up to the kinetic coefficient, which is true beyond the Gaussian theory. Thus
\begin{equation}
\lambda G({\bf k},0)=G^{\rm st}({\bf k})=\frac{1}{\tau+{\bf k}^2},
\end{equation}
where $G^{\rm st}({\bf k})$ is the static correlation function indicated by the superscript. Its inverse Fourier transforms at $d=3$ and at $\tau=0$ are proportional to $\exp(-\sqrt{\tau}|{\bf k}|)$ and $|{\bf x}|^{d-2}$, respectively, and hence confirming again Eq.~(\ref{xi}) and $\eta=0$. Such kinds of relation can be used as a check to the resultant vertex functions in dynamics as will be seen in the following sections.

Note that the Gaussian theory is in fact identical for both $\phi^4$ and $\varphi^3$ theory. This is the reason why the exponents derived in this section are also identical.

\section{\label{pert}Perturbation expansions}
Now we consider the whole model Eq.~(\ref{hp}). After Fourier transform, its associated dynamic action becomes
\begin{widetext}
\begin{equation}
{\cal L}={\cal L}_0+
\frac{1}{2}g_3'\int d\overline{{\bf k}}d\overline{{\bf k}'}d\overline{\omega}d\overline{\omega'} \tilde{\varphi}({\bf k},\omega)\varphi({\bf k}',\omega')\varphi(-{\bf k}-{\bf k}',-\omega-\omega')-\lambda \int d\overline{{\bf k}}d\overline{\omega} \tilde{\varphi}(-{\bf k},-\omega)h({\bf k},\omega), \label{L33}
\end{equation}
where ${\cal L}_0$ is given by Eq.~(\ref{L3}) and $g_3'=\lambda g_3$.
The standard method to deal with Eq.~(\ref{L33}) is to set up perturbation expansions using Feynman diagrams. One observes that $\lambda h$ plays the role of $\tilde{J}$ when comparing Eq.~(\ref{L33}) with Eq.~(\ref{z'}). Accordingly, we shall not take $\lambda h$ as an element of the expansions but rather regard it as a specific value of $\tilde{J}$.
On the basis of the Gaussian theory, let a line represent $\varphi$, a dash line $\tilde{\varphi}$, and a filled circle $-2\lambda$, then
\begin{equation}
G({\bf k},-\omega)=\begin{picture}(58,15)(10,15)
\put(12,18){\line(1,0){26}}
\multiput(41,18)(13,0){2}{\line(1,0){10}}
\end{picture},\quad
C({\bf k},\omega)=\begin{picture}(58,15)(10,15)
\put (12,18) {\line(1,0){52}}\end{picture},\quad
-\frac{1}{2}g_3'\delta\left(\sum_{i=1}^{3}{\bf k}_i\right)\delta\left(\sum_{i=1}^{3}\omega_i\right)=\begin{picture}(50,20)(10,15)
\multiput (12,18)(13,0){2}{\line(1,0){10}}
\put(35,18) {\line(3,2){20}}
\put(35,18) {\line(3,-2){20}}
\end{picture},\quad
-2\lambda =\begin{picture}(24,15)(10,15)
\multiput(12,18)(5,0){4}{\line(1,0){3}}
\put(21,18) {\circle*{3}}\end{picture}.
\end{equation}
Accordingly, to one-loop order,
\begin{eqnarray}
\Gamma_{10}({\bf k},\omega)&=&
\begin{picture}(28,18)(10,15)
\multiput(12,18)(5,0){2}{\line(1,0){3}}
\put(28,18){\circle{16}}
\end{picture}
=\frac{1}{2}g_3'(2\pi)V\delta({\bf k})\delta(\omega)\int d\overline{{\bf k}'}d\overline{\omega'}C({\bf k}',\omega')=\frac{1}{2}g_3'(2\pi)V\delta({\bf k})\delta(\omega)\int d\overline{{\bf k}'}\frac{1}{\tau+{\bf k}'^2},\label{g101}\\
\Gamma_{11}({\bf k},\omega)&=&\begin{picture}(26,20)(10,15)
\multiput(12,18)(5,0){2}{\line(1,0){3}}
\put(22,18){\line(1,0){10}}
\end{picture} +
\begin{picture}(66,23)(10,15)
\multiput(12,18)(5,0){2}{\line(1,0){3}}
\put(20,18) {\line(3,2){22}}
\put(20,18) {\line(3,-2){22}}
\multiput(64,18)(-12,8){2}{\line(-3,2){10}}
\put(64,18){\line(-3,-2){22}}
\put(64,18){\line(1,0){10}}
\end{picture}
=i\omega+\lambda(\tau+{\bf k}^2)-g_3'^2\int d\overline{{\bf k}'}d\overline{\omega'}C({\bf k}',\omega')G({\bf k}-{\bf k}',-\omega+\omega')\nonumber\\
\begin{picture}(10,6)(10,15)\end{picture}\nonumber\\
&=&i\omega+\lambda(\tau+{\bf k}^2)-g_3'^2\int d\overline{{\bf k}'}\frac{1}{\tau+{\bf k}'^2}\frac{1}{i\omega+\lambda\left[2\tau+{\bf k}'^2+({\bf k}-{\bf k}')^2\right]},\label{g111}\\
\Gamma_{20}({\bf k},\omega)&=&\begin{picture}(24,15)(10,15)
\multiput(12,18)(5,0){4}{\line(1,0){3}}
\put(21,18) {\circle*{3}}
\end{picture}+
\begin{picture}(37,18)(10,15)
\multiput(12,18)(5,0){2}{\line(1,0){3}}
\put(28,18){\circle{16}}
\multiput(36,18)(5,0){2}{\line(1,0){3}}
\end{picture}
=-2\lambda -\frac{1}{2}g_3'^2\int d\overline{{\bf k}'}d\overline{\omega'}C({\bf k}',\omega')C({\bf k}-{\bf k}',\omega-\omega')\nonumber\\
&=&-2\lambda-\lambda^3 g_3^2\int d\overline{{\bf k}'}\frac{1}{\tau+{\bf k}'^2}\frac{1}{\tau+({\bf k}-{\bf k}')^2}\frac{2\tau+{\bf k}'^2+({\bf k}-{\bf k}')^2}{\omega^2+\lambda^2\left[2\tau+{\bf k}'^2+({\bf k}-{\bf k}')^2\right]^2},\label{g201}
\end{eqnarray}
\begin{eqnarray}
\Gamma_{12}({\bf k}_1,\omega_1; {\bf k}_2,\omega_2)=\begin{picture}(23,30)(10,15)
\multiput (12,18)(5,0){2}{\line(1,0){3}}
\put(20,18) {\line(3,2){10}}
\put(20,18) {\line(3,-2){10}}
\end{picture}+
\begin{picture}(65,46)(10,15)
\multiput (12,18)(5,0){2}{\line(1,0){3}}
\put(20,18) {\line(0,3){22}}
\put(20,18) {\line(3,-2){22}}
\multiput(64,18)(-12,-8){2}{\line(-3,-2){10}}
\put(64,18){\line(0,3){22}}
\put(64,18){\line(1,0){8}}
\put(42,54.5){\line(-3,-2){22}}
\multiput(42,54.5)(12,-8){2}{\line(3,-2){10}}
\put(42,54.5){\line(0,1){8}}
\end{picture}+
\begin{picture}(65,46)(10,15)
\multiput (12,18)(5,0){2}{\line(1,0){3}}
\put(20,18) {\line(0,3){22}}
\put(20,18) {\line(3,-2){22}}
\multiput(64,18)(-12,-8){2}{\line(-3,-2){10}}
\put(64,18){\line(0,3){22}}
\put(64,18){\line(1,0){8}}
\put(42,54.5){\line(3,-2){22}}
\multiput(42,54.5)(-12,-8){2}{\line(-3,-2){10}}
\put(42,54.5){\line(0,1){8}}
\end{picture}\nonumber\\
\begin{picture}(10,6)(10,15)\end{picture}\nonumber\\
=g_3'+g_3'^3\int d\overline{{\bf k}}d\overline{\omega}\left[2C({\bf k},\omega)G({\bf k}_1-{\bf k},\omega-\omega_1)G({\bf k}_2+{\bf k},\omega+\omega_2)+C({\bf k},\omega)G({\bf k}_3-{\bf k},\omega-\omega_3)G({\bf k}-{\bf k}_2,\omega_2-\omega)\right]\nonumber\\
=g_3'+g_3'^3\int d\overline{{\bf k}}\frac{1}{\tau+{\bf k}^2}\left\{\frac{2}{i\omega_1+\lambda\left[2\tau+{\bf k}^2+({\bf k}_1-{\bf k})^2\right]}\frac{1}{-i\omega_2+\lambda\left[2\tau+{\bf k}^2+({\bf k}_2+{\bf k})^2\right]}\right.+\begin{picture}(99,6)\end{picture}\nonumber\\
\left.\frac{1}{i\omega_1+\lambda\left[2\tau+({\bf k}-{\bf k}_2)^2+({\bf k}_3-{\bf k})^2\right]}
\left[\frac{1}{i\omega_3+\lambda\left[2\tau+{\bf k}^2+({\bf k}_3-{\bf k})^2\right]}+\frac{1}{-i\omega_2+
\lambda\left[2\tau+{\bf k}^2+({\bf k}-{\bf k}_2)^2\right]}\right]\right\}, \begin{picture}(13,6)\end{picture}\label{g121}
\end{eqnarray}
where the short lines indicate the vertices for clarity and do not enter the expressions of the vertex functions, ${\bf k}_3={\bf k}_1+{\bf k}_2$, and $\omega_3=\omega_1+\omega_2$. One can check that when all external frequencies become zero, the vertex functions indeed recover their static counterparts. For example, the integral in Eq.~(\ref{g111}) remains unchanged after a replacement ${\bf k}'\rightarrow{\bf k}-{\bf k}'$. Adding these two integrals up results in
\begin{equation}
\Gamma_{11}({\bf k},0)=\lambda(\tau+{\bf k}^2)-\frac{1}{2}\lambda g_3^2\int d\overline{{\bf k}'}\frac{1}{\tau+{\bf k}'^2}\frac{1}{\tau+({\bf k}-{\bf k}')^2},\label{g1110}
\end{equation}
which is $\lambda\Gamma_{2}^{\rm st}({\bf k})$. Similarly, one can show that
\begin{equation}
\Gamma_{12}({\bf k}_i,0)=\lambda g_3-\lambda g_3^3\int d\overline{{\bf k}}\frac{1}{\tau+{\bf k}^2}\frac{1}{\tau+({\bf k}_1-{\bf k})^2}\frac{1}{\tau+({\bf k}_2+{\bf k})^2},\label{g1210}
\end{equation}
\end{widetext}
which is just $\lambda\Gamma_{3}^{\rm st}({\bf k}_i)$. Moreover, one can also check that the fluctuation-dissipation theorem Eq.~(\ref{fdt20}) does obey to this order by a similar manipulation.

An effect of the interaction is the shift of the instability point. It is defined by
\begin{subequations}
\begin{eqnarray}
\Gamma_{11}({\bf 0},0)=0,\label{taus}\\
\Gamma_{10}({\bf 0},0)=\lambda h_s,\label{hs}
\end{eqnarray}
\end{subequations}
which are
\begin{subequations}
\label{tauhs1}
\begin{eqnarray}
\tau_s=\frac{1}{2}g_3^2\int d\overline{{\bf k}}\frac{1}{(\tau_s+{\bf k}^2)^2},\label{taus1}\\
h_s=\frac{1}{2}g_3\int d\overline{{\bf k}}\frac{1}{\tau_s+{\bf k}^2}\label{hs1}
\end{eqnarray}
\end{subequations}
for a uniform external field $H$ and hence $h$ and $M$. Eqs.~(\ref{taus}) and (\ref{taus1}) determine the point at which the static susceptibility $\chi$ diverges (see below). Eqs.~(\ref{hs}) and (\ref{hs1}) specifies the field at $\tau_s$ for
\begin{equation}
m(\tau_s,h_s)=\langle\varphi(\tau_s,h_s)\rangle=0.\label{mths}
\end{equation}
It can be shown explicitly that Eq.~(\ref{tauhs1}) is exactly the instability point of the one-loop $\varphi^3$ free energy at $m=0$ but the corresponding $\phi^4$ free energy has a vanishingly small barrier in accordance with what observed near the spinodal point in non-mean-field systems. Accordingly, this instability point has been advocated as a new definition of a general spinodal point \cite{zhongjcp}. As the mean-field spinodal given by Eq.~(\ref{mfs}) is $\tau_s=0$, one can replace $\tau_s$ in the integrand in Eq.~(\ref{taus1}) by its mean-field value in a perturbation expansion. So,
\begin{subequations}
\label{tauhs10}
\begin{eqnarray}
\tau_s=\frac{1}{2}g_3^2\int d\overline{{\bf k}}\frac{1}{{\bf k}^4},\label{taus10}\\
h_s=\frac{1}{2}g_3\int d\overline{{\bf k}}\frac{1}{{\bf k}^2}.\label{hs10}
\end{eqnarray}
\end{subequations}
The first expression is infrared divergent for $d\leq4$ and the second for $d\leq2$ because they diverge when $|{\bf k}|\rightarrow0$ in these dimensions. Note that the momentum integral is cutoff at $\Lambda$.

The infrared divergences in Eq.~(\ref{tauhs10}) affect the position of the instability point which is a property of a specific system. They can in fact be subtracted by mass renormalization as we shall do shortly. However, these infrared divergences which stem from the small momentum behavior at the instability point persist and in fact plague the whole perturbation expansion below an upper critical dimensions.

Equation~(\ref{taus10}) can be employed to write the static susceptibility $\chi$ as
\begin{equation}
\chi^{-1}=\Gamma_{11}({\bf 0},0)/\lambda=\tau_r\left[1-\frac{1}{2}g_3^2\int d\overline{{\bf k}}\frac{\tau_r+2{\bf k}^2}{(\tau_r+{\bf k}^2)^2{\bf k}^4}\right]\label{xmr}
\end{equation}
since the correction incurred is of higher order in $g_3$, where $\tau_r=\tau-\tau_s$. Note that the term proportional to $\tau_r$ in the integral equals zero when $\tau_r=0$, one sees then one can again set $\tau_r$ in the integrand to its mean-field value if $d>6$ because the integral is then finite. As a result, $\chi$ is again proportional to $\tau_r^{-\gamma}$ with the mean-field result $\gamma=1$. However, if $d\leq6$, the integral has an infrared divergence at the instability point and the mean-field result becomes problematic. In fact, in this case, the integral is dominated by small $|{\bf k}|$s and so can be well approximated by
\begin{equation}
\chi^{-1}=\tau_r\left[1-\frac{\Gamma\left(3-\frac{d}{2}\right)}{(4\pi)^{d/2}(d-4)}g_3^2 \tau_r^{\frac{d}{2}-3}\right] \label{xint}
\end{equation}
by extending the momentum cutoff to infinity and using Eqs.~(\ref{iqqn}), (\ref{nd}), and (\ref{gz1}), where $\Gamma$ here is the Euler Gamma function. One sees from Eq.~(\ref{xint}) that the second term in the brackets can become arbitrarily larger than 1, the first term, for sufficiently small $\tau_r$ and the perturbation expansion breaks down.
This therefore identifies the upper critical dimension $d_c=6$ above which the mean-field theory of the $\varphi^3$ model is valid. Accordingly, the spinodal curve and also the mean-field spinodal fixed point \cite{Gunton78} which build on the mean-field theory all become valid above $d_c$ in agreement with previous mean-field analyses \cite{Gunton78,Klein83}.

Moreover, one can even introduce a Ginzburg effective temperature $\tau_G$ to write Eq.~(\ref{xint}) as
\begin{equation}
\chi^{-1}=\tau_r\left[1-\left(\frac{\tau_r}{\tau_G}\right)^{\frac{d-6}{2}}\right]\label{taug}
\end{equation}
similar to the critical phenomena \cite{Ginzburg,Amit74}. Accordingly, for $d>6$, sufficiently small $\tau_r$, i.e., sufficiently close to the instability point can always make the second term in the square brackets smaller than the first one and the mean-field theory is valid. However, for $d<6$, this is only true for $|\tau_r|\gg\tau_G$, which is the Ginzburg criterion \cite{Binder84}. This therefore defines an unstable region similar to the critical region within which fluctuations dominate and the mean-field theory fails. Note that, from Eq.~(\ref{xint}), this is always true in $d=4$. This special dimension would have become 2 \cite{Binder84} if the term proportional to $\tau_r$ in the integral in Eq.~(\ref{xmr}) had been neglected.

We shall show in the following (Sec.~\ref{count}) that higher order corrections also confirm this differentiation of spatial dimensions. Note that our analysis here is completely parallel to that in critical phenomena \cite{Justin,Brezin,amitb}. Therefore, the natural solution to the infrared divergence problem below $d_c$ is the RG theory.

\section{\label{rgt}Renormalization-group theory}
We have shown in the previous sections that an FOPT can be mapped to the instability point of a $\varphi^3$ model and the long wavelength fluctuations near the point result in strong infrared divergences below $d_c$ in close analogy to the critical phenomena. In this section, we shall apply the RG theory in the field-theoretic formulation \cite{Brezin,Janssen79,Janssen,Justin,amitb,Vasilev,Tauber} to the $\varphi^3$ theory. The RG technique is an effective method for treating both infrared and ultraviolet singularities \cite{Vasilev}. The usual field-theoretic formulation often concerns with the latter. We shall discuss their relation by applying the usual approach to the present theory and formulate the RG theory for the transition depicted in Fig.~\ref{mff}(c), an RG for a massless theory, i.e., the RG for the $\varphi^3$ theory at the point where Eq.~(\ref{taus}) is satisfied similar to the critical massless theory.

\subsection{\label{dimensions}Canonical dimensions}
\begin{table*}[ht]
\caption{\label{dim}Canonical dimensions}
\begin{ruledtabular}
\begin{tabular}{cccccccccccccc}
$|{\bf x}|$ & $|{\bf k}|$ & $\lambda t$ & $\tau$ & $\varphi(x)/M$& $\tilde{\varphi}(x)$& $g$ & $g_3$  & $h/H/\lambda\tilde{J}$&$\lambda J$ & $\lambda\varphi({\bf k},\omega)$&$\lambda\tilde{\varphi}({\bf k},\omega)$&$\lambda^{n+n'-1}G_{nn'}(\{{\bf k}_i\},\{\omega_i\})$&$\lambda^{-1}\Gamma_{n'n}(\{{\bf k}_i\},\{\omega_i\})\equiv d_{n'n}$\\
\hline
$-1$ & 1& $-2$  &2 & $\frac{d-2}{2}$ & $\frac{d+2}{2}$ &$4-d$&$\frac{6-d}{2}$& $\frac{d+2}{2}$&$\frac{d+6}{2}$&$-\frac{d+6}{2}$&$-\frac{d+2}{2}$&$\frac{2-n'-n}{2}d-n'-3n+2$&$\frac{2-n'-n}{2}d+n-n'+2$      \\
\end{tabular}
\end{ruledtabular}
\end{table*}
To start with, we determine the canonical dimensions of various quantities \cite{Justin,amitb}. These can be obtained by a na\"{\i}ve dimensional analysis in which ${\cal H}$, ${\cal H}_3$, and ${\cal L}$ are regarded as pure numbers as they appear in exponentials and all other quantities have the dimension of length, which is taken to be $-1$, i.e., $[|{\bf x}|]=-1$, where the square brackets here denote the canonical dimension of the quantity in them. So, $[|{\bf k}|]=1$.  Eq.~(\ref{L3}) then leads to $[\lambda t]=-2$ and $[\tau]=2$ as $\partial/\partial t/\lambda$ and $\tau$ play a similar role to $\nabla^2$. Consequently, $[\tilde{\varphi}]=(d+2)/2$ and $[\varphi]=(d-2)/2$ from Eq.~(\ref{L3}) and hence $[\lambda J]=(d+6)/2$ and $[\lambda\tilde{J}]=(d+2)/2$ from Eq.~(\ref{zjj}), where the $\lambda$ factors arise from the time integral. As a result, $[g_3]=(6-d)/2$, $[h]=[\lambda\tilde{J}]=(d+2)/2$ from Eqs.~(\ref{Lj}) and (\ref{hp}). Similarly, one finds $[r]=2$, $[\phi]=(d-2)/2$, $[\tilde{\phi}]=(d+2)/2$, $[H]=(d+2)/2$, and $[g]=4-d$ for the dynamic $\phi^4$ model. Note that $[g_3]=[g]+[M]$ since $[M]=[\phi]$. These dimensions are identical with their static counterparts as can be readily seen from Eqs.~(\ref{H}) and (\ref{hp}). In addition, from Eq.~(\ref{ifou}), $[\lambda\varphi({\bf k},\omega)]=-(d+6)/2$ and $[\lambda\tilde{\varphi}({\bf k},\omega)]=-(d+2)/2$.

Accordingly, one finds from Eqs.~(\ref{gcd}), (\ref{ga}), and (\ref{g}) that $[G_{nn'}(\{x_i\})]=n[\phi]+n'[\tilde{\phi}]=(n+n')d/2+n'-n$ and $[\lambda^{n'+n}\Gamma_{n'n}(\{x_i\})]=n[\lambda J]+n'[\lambda\tilde{J}]=(n+n')d/2+n'+3n$, the latter can also be obtained by attaching back the amputated legs of $\Gamma_{n'n}(\{x_i\})$ with integrations and relating it to its corresponding response function. We have used $\{x_i\}$ to represent the set of all the spatial and temporal arguments. Then, from the Fourier transforms defined in Eq.~(\ref{gk2k}) and a similar definition for the vertex function, one obtains $[\lambda^{n+n'}G_{nn'}(\{{\bf k}_i\},\{\omega_i\})]=[G_{nn'}(\{x_i\})]-(n+n')(d+2)=-(n'+n)d/2-n'-3n=n[\lambda\varphi({\bf k},\omega)]+n'[\lambda\tilde{\varphi}({\bf k},\omega)]$ and $[\Gamma_{n'n}(\{{\bf k}_i\},\{\omega_i\})]=[\lambda^{n+n'}\Gamma_{n'n}(\{x_i\})]-(n+n')(d+2)=-(n'+n)d/2+n-n'$. The response functions and vertex functions we have primarily used in Secs.~\ref{gauss} and \ref{pert} and shall be used below have, however, an extracted delta function factor of dimension $-d-2$ arising from the global momentum and frequency conservations due to space and time translation invariances as seen in Eq.~(\ref{gk2k}). Consequently, their dimensions change by this amount and are listed in Table~\ref{dim} using the same symbols but with one power of $\lambda$ less due to the same reason.

\subsection{\label{count}Power counting and divergences}
We now consider the relation of the infrared divergences to the ultraviolet ones and show that the consequence of the one-loop analysis in the perturbation expansion above survives to all orders. In particular, the $\varphi^3$ theory is non-renormalizable and thus the mean-field results are valid for $d>d_c$ but super-renormalizable for $d<d_c$ in which the mean-field theory fails and only the cubic interaction is relevant if a massless theory exists.

Take Eq.~(\ref{hs1}) as an example. Assume that a mass renormalization has been performed such that all $\tau$s in the response functions have been replaced with $\tau_r$s, which vanish at the instability point. The integral diverges for $d<2$ because it behaves as $|{\bf k}|^{d-2}$ when $\tau_r=0$ as can be readily seen from Eq.~(\ref{hs10}). If we change the variable to ${\bf k}/\sqrt{\tau_r}$, we have
\begin{equation}
h_s=\frac{1}{2}g_3\tau_r^{d/2-1}\int d\overline{{\bf k}}\frac{1}{1+{\bf k}^2},\label{hs1i}
\end{equation}
where the cutoff now becomes $\Lambda/\sqrt{\tau_r}$. For $d>2$, the integral is dominated by large $|{\bf k}|$s and diverges as $(\Lambda/\sqrt{\tau_r})^{d-2}$. One sees, however, that in this case the $\tau_r$ factor just cancels the one outside the integral and $h_s\sim\Lambda^{d-2}$, a constant. So, no divergence appears as should be. However, for $d<2$, the integral is now dominated by small $|{\bf k}|$s and can be integrated using Eq.~(\ref{iqn}) when $\tau_r\rightarrow0$, leading to
\begin{equation}
h_s=\frac{1}{2(4\pi)^{d/2}}g_3\Gamma\left(1-\frac{d}{2}\right)\tau_r^{d/2-1}.\label{hs1i1}
\end{equation}
This result coincides with what would arise from a direct integration of Eq.~(\ref{hs1}) for small $\tau_r$ and thus describes the infrared divergence at the instability point correctly. However, as the new cutoff is now infinity at the instability point, the integral can be directly integrated without the approximated argument that leads to Eq.~(\ref{xint}).
Accordingly, we can study directly the ultraviolet behavior of large $\Lambda$ in both cases. This enables us to employ the machinery of the field-theoretic RG theory for ultraviolet divergences. From the point of view of statistical physics, this RG procedure singles those universal properties out of the microscopic details specified by $\Lambda$.

Before extending the analysis to all orders of the perturbation expansion, we first consider in general the asymptotic dependence of $\Gamma_{n'n}(\{{\bf k}_i\},\{\omega_i\})$ on $\Lambda$, i.e., its large momentum behavior. We do this by only considering primitive dependence and neglecting subintegrations \cite{Justin,amitb} as an illustration. At the $N$th order, a graph of $L$ loops contributing to $\Gamma_{n'n}$ contains $L$ integrations, $N$ vertices each emitting 3 lines, $[3N-(n+n')]/2$ internal lines among which $N-n'$ are response lines each having a ${\bf k}^{-2}$ factor and the remaining $(N-n+n')/2$ correlation lines each having a ${\bf k}^{-4}$ factor for large $|{\bf k}|$s. Consequently, the graph has a total ${\bf k}$ factor of
\begin{eqnarray}
\delta_{n'n}&=&L(d+2)-2(N-n')-4\times\frac{1}{2}(N-n+n')\nonumber\\
&=&L(d+2)-4N+2n\label{dnn}
\end{eqnarray}
powers and thus is proportional to $\Lambda^{\delta_{n'n}}$. Note that the frequency integrations are also relevant to momentum factors as can be checked in Eqs.~(\ref{g101}) to (\ref{g121}). The $L$ internal integrations comes from the total internal lines less the number of momentum and frequency conservations at each vertex except for one for the overall conservation as may be seen in Eq.~(\ref{gk2k}). So,
\begin{equation}
L=\frac{1}{2}[3N-(n+n')]-N+1=\frac{1}{2}(N-n-n')+1.\label{Lnn}
\end{equation}
Combining Eqs.~(\ref{dnn}) and (\ref{Lnn}) leads to
\begin{eqnarray}
\delta_{n'n}&=&\frac{1}{2}(2-n-n')d+n-n'+2+\frac{1}{2}(d-6)N\nonumber\\
&=&d_{n'n}-N[g_3],\label{dnn1}
\end{eqnarray}
where we have use the dimensions of the two quantities given in Table~\ref{dim}. The last expression is apparent as the dependence on $\Lambda$ must be deducted from the total dimension the contribution from the coupling constant $g_3$. All these results may also be checked with Eqs.~(\ref{g101}) to (\ref{g121}).

Equation~(\ref{dnn1}) has one overall factor depending only on $d$ and on $n$ and $n'$ of the vertex function and the other depending linearly both on the order of the perturbation expansion and on the dimension of the coupling constant. In fact, as the first factor does not depend on the coupling, it can be shown to be the dimension of the vertex function in the Gaussian theory. One concludes therefore that if the coupling constant is dimensionless, the cutoff-dependence of the vertex functions will be independent of the order in the perturbation theory. So will the resultant primitive divergences of all the vertex functions when $\Lambda\rightarrow\infty$. This identifies again a critical dimension $d_c=6$ at which the coupling constant is dimensionless for the $\varphi^3$ model considered. For $d>d_c$, the ultraviolet divergences of the graphs of a vertex function increase their number with the order of perturbation expansion. As a result, they cannot be absorbed into a finite number of renormalized parameters which contain the strong $\Lambda$ or system dependence. The theory in such a case is then termed non-renormalizable. On the other hand, for $d<d_c$, the degree of the primitive divergences decreases as one goes to higher orders in the perturbation theory and the theory is then termed super-renormalizable, while a theory in $d=d_c$ is renormalizable \cite{Justin,Brezin,amitb}.

Now we relate this asymptotic dependence to the infrared behavior of the graph which contributes $D_{n'n}$ to $\Gamma_{n'n}$. For simplicity let us set all its external momenta and frequencies zero and change again its integration variables by
\begin{equation}
{\bf k}'={\bf k}/\sqrt{\tau_r}, \qquad \omega'=\omega/\tau_r.
\end{equation}
As a result,
\begin{equation}
D_{n'n}=\tau_r^{\delta_{n'n}/2}D'_{n'n}.
\end{equation}
The remaining integral in $D'_{n'n}$ has no infrared divergences at $\tau_r\rightarrow0$ as can be seen from Eq.~(\ref{hs1i}). However, when $d>d_c$, i.e., the interaction is non-renormalizable, for sufficiently high orders in the perturbation expansion, $D'_{n'n}$ has ultraviolet divergences proportional to $[\Lambda/\sqrt{\tau_r}]^{\delta_{n'n}}$. Yet, similar to our example, the two factors of $\tau_r$ cancel exactly, leaving only finite numbers in every order in the perturbation. As a result, the mean-field behavior survives. For those vertex functions with large $n'$ and $n$ (for $d>2$), there may exist some lower order diagrams that have $\delta_{n'n}<0$ from Eq.~(\ref{dnn1}) and are thus ultraviolet convergent. As $\delta_{n'n}$ increases with the order $N$, the leading infrared divergence comes from the lowest one-loop term with only vertices having the smallest $d_{n'n}$ \cite{Justin}.

When $d<d_c$ on the other hand, the theory is super-renormalizable and only a finite number of diagrams have ultraviolet divergences. These divergences can then be incorporated into a redefinition of the model parameters such that the theory built on these redefined parameters converges at large momenta. This means that the large scale behavior of the theory does not depend on its microscopic details prescribed by $\Lambda$ and thus universality ensues.

Note that, for $d<d_c$, except those ultraviolet divergent diagrams, the others are superficially convergent by power counting and the infrared divergences are thus given by $\tau_r^{\delta_{n'n}/2}$ or may be even more singular when $\tau_r\rightarrow0$. Moreover, as Eq.~(\ref{dnn1}) shows, these singularities increase without bound with the order of perturbation. However, we now demonstrate that the most infrared divergent terms order by order in the perturbation theory are described by the $\varphi^3$ theory \cite{Justin,Brezin,amitb}. All higher order couplings are irrelevant as far as the infrared behavior is concerned.

Consider, for example, a theory that contains $\iota$ types of interaction vertices each emitting one $\tilde{\varphi}$ line and $n_{\iota}-1$ $\varphi$ lines. For such a theory, Eqs.~(\ref{dnn}) and (\ref{Lnn}) change to
\begin{eqnarray}
\delta'_{n'n}&=&L'(d+2)-2\sum_{\iota}N_{\iota}n_{\iota}+2N+2n,\label{dnnp}\\
L'&=&\frac{1}{2}\left[\sum_\iota N_{\iota}n_{\iota}-(n+n')\right]-N+1,\label{Lnnp}
\end{eqnarray}
and so,
\begin{eqnarray}
\delta'_{n'n}&=&\frac{1}{2}(2-n-n')d+n-n'+2\nonumber\\
&&+\sum_{\iota}N_{\iota}\left(\frac{1}{2}n_{\iota}d-n_{\iota}-d\right)\nonumber\\
&=&d_{n'n}-\sum_{\iota}N_{\iota}[g_{\iota}],\label{dnn1p}
\end{eqnarray}
where $N=\sum_{\iota}N_{\iota}$ and $[g_{\iota}]=n_{\iota}+d-n_{\iota}d/2$ is the dimension of the coupling with $n_{\iota}$ fields with $g_4\equiv g$. Now one sees that $-[g_{\iota}]= n_{\iota}(d-2)/2-d$ increases with $n_{\iota}$ for $d>2$. The minimum $n_{\iota}$ thus gives minimum $n_{\iota}(d-2)/2-d$ and hence minimum $\delta'_{n'n}$, which is the most negative value and thus characterizes the most infrared divergent diagrams in each order of the perturbation expansion. In the case of the critical phenomena of the Ising model, its order parameter has an inversion symmetry. The smallest $n_{\iota}$ is thus 4 and thus the $\phi^4$ model reproduces the sum of the most infrared divergent contributions order by order in a mean-field expansion below its critical dimension $d_c=2n_{\iota}/(n_{\iota}-2)=4$ at which $[g_{\iota}]=0$ \cite{Justin,Brezin}. In systems with FOPTs, as their symmetries have already been broken, the smallest $n_{\iota}$ is thus 3 and thus all other terms including terms with more spatial derivatives which give rise to ${\bf k}$ factors in the numerator of the momentum integrals are infrared irrelevant for $d<6$. At $d=6$, mean-field behavior is modified by logarithmic corrections as in critical phenomena.

\subsection{\label{rgsrc}RG scheme and renormalization constants}
We have seen that for $d<6$ the infrared singularities at the instability points of an FOPT can also be described by a $\varphi^3$ theory. In these dimensions, the ultraviolet divergences of the theory are super-renormalizable and can thus be absorbed in a few parameters by renormalization. There are usually two perturbation RG schemes to deal with these divergences. One uses the $\varepsilon=6-d$ expansion and the other works directly in a fixed dimension. We shall adopt the first one. We shall only study the massless theory as mentioned. This is because we have been focusing on the FOPT described by Fig.~\ref{mff}(c). Although the massless theory has problems associated with infrared divergences and cannot yield results such as universal amplitude ratios \cite{Symanzik,Parisi,Bergere,Bagnuls,Dohm}, it can nevertheless give exponents that we concern here. We shall apply the technique of dimensional regulation to the ultraviolet divergent integrals and utilize the minimal RG method to subtract and just subtract the resultant dimensional poles \cite{hooft}. In fact, this technique of analytically continuation in the number of space dimensions and pushing $\Lambda$ to infinity has been argued to be a more natural approach to the critical phenomena and similarly to the `unstable' phenomena here as one can then focus on the infrared behavior alone without resorting to the ultraviolet divergences in a theory with a finite cutoff \cite{Amit76}. The RG method of minimal subtraction has an advantage of decoupling dynamics from statics without having to choose deliberately renormalization conditions \cite{de}. As a result, the static renormalization factors can be chosen to be identical to those of the corresponding equilibrium model. This can be used as a self-check and can also simplify manipulations.

As we shall use the $\varepsilon$ expansion, we shall renormalize the theory at $d_c=6$. The first task is to identify those vertex functions that are superficially divergent. They are given by $\delta_{n'n}\geq0$ at $d_c$, which, at this dimension, is just $d_{n'n}$ of the pertinent vertex functions. From Table~\ref{dim}, one finds these vertex functions are $\Gamma_{10}$, which is quartic divergent, $\Gamma_{11}$, which is quadratic divergent, and $\Gamma_{20}$ and $\Gamma_{12}$, both of which are logarithmic divergent. So, one should in principle introduce four renormalization constants to absorb the divergences. However, they are not all independent. $\Gamma_{20}$ is related to $\Gamma_{11}$ by Eq.~(\ref{fdt20}). Consequently, we need introduce only three renormalization constants. Of course, we have to perform subtractions or mass renormalizations to deal with the quadratic and quartic divergences of $\Gamma_{10}$ and $\Gamma_{11}$. However, as we shall study the massless theory and utilize the technique of dimensional regulation, these subtractions become zero because massless integrals such as Eq.~(\ref{tauhs10}) are zero in the dimensional regulation. This does not mean, of course, that the instability point is not shifted. Note that we shall not consider composite operators and the free energy.

Although we shall work with the massless theory, we shall consider an external field that drives the transition as Fig.~\ref{mff}(c) shows. Accordingly, we rewrite our relevant action as
\begin{equation}
{\cal L} = \int d{\bf x}dt \tilde {\varphi } \left[ \frac{d
\varphi}{dt} -\lambda \left( \nabla ^2\varphi - \frac{1}{2}g_{3} \varphi ^2 +
h + \tilde {\varphi} \right)\right] , \label{actions}
\end{equation}
where we have neglected $\tau_s $ and $h_s$ as they are zero in the dimensional regulation. The renormalizability of this massless theory for $d\leq6$ means that by choosing the renormalization factors $Z(u_R,\varepsilon)$s as \begin{equation}
\begin{array}{c}
\varphi = Z_{\varphi} ^{1 / 2} \varphi_R ,\qquad \tilde {\varphi } =
Z_{\tilde {\varphi }}^{1 / 2} \tilde {\varphi }_R, \qquad \lambda = Z_{\lambda}\lambda_R , \\
u = Z_u u_R, \qquad u = g_{3} N_d^{1 / 2} \mu ^{-\varepsilon / 2},
\end{array}
\label{z}
\end{equation}
the renormalized $G_{nn'R}$ and $\Gamma_{n'nR}$ as functions of the renormalized functions $\lambda_R$ and $u_R$ satisfying
\begin{eqnarray}
G_{nn'}^c=Z_{\varphi}^{n/2}Z_{\tilde {\varphi }}^{n' / 2}G_{nn'R},\label{gnr}\\
\Gamma_{n'n}=Z_{\tilde {\varphi }}^{-n' / 2}Z_{\varphi}^{-n/2}\Gamma_{n'nR}\label{gar}
\end{eqnarray}
from their definitions are finite at every order in an expansion in $u_R$ as $\varepsilon\rightarrow0$, where $\mu$ is an arbitrary momentum scale, $N_d$ is given by Eq.~(\ref{nd}), and the subscripts $R$ denote renormalized quantities. As the fluctuation-dissipation theorem must hold in both the unrenormalized and the renormalized forms, one finds from Eqs.~(\ref{fdt20}), (\ref{z}), and (\ref{gar})
\begin{equation}
Z_{\lambda}=Z_{\varphi} ^{1 /2} Z_{\tilde {\varphi }}^{ - 1 / 2}.\label{zl1}
\end{equation}
This confirms that only three $Z$s are needed.

We now compute these factors to one-loop order using the method of minimal subtraction and show that they indeed make the vertex functions finite.

Firstly, we choose Eq.~(\ref{g1110}) to fixed $Z_{\varphi}$. Note that all $\tau$ must be set to zero when we refer to the equations in Sec.~\ref{pert} in the following sections as we are considering a massless theory. Using Eq.~(\ref{ikpk}) and noting that $\Gamma(2-d/2)$ contributes an $\varepsilon$ pole from Eq.~(\ref{gep}) and hence the other $\Gamma$ functions can be evaluated directly at $d=6$, one finds
\begin{equation}
\Gamma_{11}({\bf k},0)=\lambda{\bf k}^2+\frac{1}{6\varepsilon}\lambda g_3^2N_d\left({\bf k}^2\right)^{\frac{d}{2}-2}\left[1+O(\varepsilon)\right].\label{g111e}
\end{equation}
As mentioned in Sec.~\ref{pert}, $\Gamma_{11}({\bf k},0)=\lambda\Gamma_2^{\rm st}({\bf k})$. So, using the definition of $u$ in Eq.~(\ref{z}), we have
\begin{equation}
\Gamma_{2}^{\rm st}({\bf k})={\bf k}^2\left\{1+\frac{1}{6\varepsilon}u^2\left[1+O(\varepsilon)\right]\right\},\label{g2e}
\end{equation}
where we have expanded $(|{\bf k}|/\mu)^{-\varepsilon}=1-\varepsilon\ln(|{\bf k}|/\mu)+O(\varepsilon^2)$ and neglected all the terms that add to order $O(\varepsilon)$ in Eq.~(\ref{g2e}) consistently. One sees that the neglected $\ln(|{\bf k}|/\mu)$ can never overpower the overall ${\bf k}^2$ factored out in Eq.~(\ref{g2e}) in the infrared limit and thus one has a renormalized massless theory in every order in $u$ such that $\Gamma_{11}({\bf 0},0)=0$. One sees also that the residue of the pole is independent of the external momentum. In fact, this is a general feature by which the residue of the pole of highest order in a graph must be independent of the external momenta and the momentum dependence of the poles of lower order must be canceled for any set of external momenta \cite{amitb}. This offers a check to the calculations.

According to Eq.~(\ref{gar}), one can choose $Z_{\varphi}$ in the method of minimal subtraction to subtract and just subtract the pole in Eq.~(\ref{g2e}) such that
\begin{equation}
\Gamma_{2R}^{\rm st}({\bf k},\mu)=Z_{\varphi}\Gamma_{2}^{\rm st}({\bf k}) \label{g2r}
\end{equation}
is finite. This leads to
\begin{equation}
Z_\varphi = 1 - \frac{1}{6\varepsilon}u^2.\label{zphi}
\end{equation}

Next, we choose Eq.~(\ref{g111}) with ${\bf k}={\bf 0}$ to fix $Z_{\tilde{\varphi}}$. Using Eq.~(\ref{iqqn}) formally and factoring the $\varepsilon$ pole out, one finds
\begin{eqnarray}
\Gamma_{11}({\bf 0},\omega)&=&i\omega+\frac{1}{2\varepsilon}\lambda g_3^2N_d \left(\frac{i\omega}{2\lambda}\right)^{\frac{d}{2}-2}\left[1+O(\varepsilon^2)\right]\nonumber\\
&=&i\omega\left\{1+\frac{1}{4\varepsilon}u^2\left[1+O(1)\right]\right\},\label{g11e}
\end{eqnarray}
where the definition of $u$ in Eq.~(\ref{z}) has been used and powers of $\ln(i\omega/2\lambda\mu^2)$ has also been neglected. One observes similar features to the previous case here.

According to Eq.~(\ref{gar}), we then choose
\begin{equation}
(Z_{\varphi}Z_{\tilde{\varphi}})^{1/2}=1-\frac{1}{4\varepsilon}u^2\label{zz2}
\end{equation}
to subtract the pole in Eq.~(\ref{g11e}) and obtain
\begin{equation}
Z_{\tilde {\varphi }} = 1 -\frac{1}{3\varepsilon}u^2\label{ztphi}
\end{equation}
using Eq.~(\ref{zphi}).

Equation~(\ref{ztphi}) can be checked using the usual method. It uses
\begin{eqnarray}
\left.\frac{\partial\Gamma_{11}({\bf k},\omega)}{\partial(i\omega)}\right|_{\omega=0} &=& 1+g_3^2I_1({\bf k})\begin{picture}(80,12)\end{picture}\nonumber\\
&=&1+\frac{1}{4\varepsilon}u^2\left[\frac{|{\bf k}|}{\mu}\right]^{-\varepsilon} \left[1+O(\varepsilon)\right]
\end{eqnarray}
from Eqs.~(\ref{i1}) and (\ref{i1e}). This leads therefore again to Eqs.~(\ref{zz2}) and (\ref{ztphi}). One can also use directly
\begin{eqnarray}
\Gamma_{20}({\bf k},0)&=&-2\lambda-\lambda g_3^2I_2({\bf k})\begin{picture}(99,12)\end{picture}\nonumber\\
&=&-2\lambda\left\{1+ \frac{1}{4\varepsilon}u^2 \left[\frac{|{\bf k}|}{\mu}\right]^{-\varepsilon}\left[1+O(\varepsilon)\right]\right\}
\end{eqnarray}
from Eqs.~(\ref{i2}) and (\ref{i2e}). So, Eq.~(\ref{gar}) results in
\begin{eqnarray}
\Gamma_{20R}({\bf k},0,\mu)&=&Z_{\tilde{\varphi}}\Gamma_{20}({\bf k},0)\begin{picture}(116,12)\end{picture}\nonumber\\
&=&-2\lambda_RZ_{\lambda}Z_{\tilde{\varphi}}\left\{1+ \frac{1}{4\varepsilon}u^2\left[1+O(\varepsilon)\right] \right\}
\end{eqnarray}
with the help of Eq.~(\ref{z}). Therefore, one has
\begin{equation}
Z_{\lambda}Z_{\tilde{\varphi}}=1-\frac{1}{4\varepsilon}u^2 =(Z_{\varphi}Z_{\tilde{\varphi}})^{1/2},\label{zpl}
\end{equation}
which recovers Eq.~(\ref{zl1}) as it should be. Of course, because the latter is true, Eq.~(\ref{zpl}) gives rise again consistently to Eq.~(\ref{ztphi}) using Eq.~(\ref{zphi}).

Finally, we utilize Eq.~(\ref{g1210}) to determine $Z_u$. One has
\begin{eqnarray}
\Gamma_{3R}^{\rm st}({\bf k},\mu)&=&\Gamma_{12R}({\bf k},0,\mu)/\lambda_R=Z_{\varphi}^{3/2}\Gamma_{12R}({\bf k},0,\mu)/\lambda_R\nonumber\\
&=&Z_{\varphi}^{3/2}\left[g_3+g_3^3I_3({\bf k}_1,{\bf k}_2)\right]\nonumber\\
&=&Z_{\varphi}^{3/2}Z_uu_RN_d^{-1/2}\mu^{\varepsilon/2}\left\{ 1+\frac{1}{\varepsilon}u^2\left[1+O(\varepsilon)\right]\right\}\nonumber\\
\end{eqnarray}
from Eqs.~(\ref{gar}), (\ref{i3}), (\ref{i3e}), and (\ref{z}). So,
\begin{equation}
Z_{\varphi}^{3/2}Z_u=1-\frac{1}{\varepsilon}u^2,
\end{equation}
which leads to
\begin{equation}
Z_u= 1 - \frac{3}{4\varepsilon}u^2,\label{zu}
\end{equation}
using Eq.~(\ref{zphi}).

However, all the $Z$ factors must be series in $u_R$ instead of $u$ itself. Using Eqs.~(\ref{z}) and (\ref{zu}), one obtains
\begin{equation}
u=Z_uu_R=u_R-\frac{3}{4\varepsilon}u_R^3\label{uur}
\end{equation}
by iteration. Therefore, to the one-loop order, all $u^2$ can simply be replaced by $u_R^2$ and we finally have
\begin{equation}
Z_\varphi = 1 - \frac{1}{6\varepsilon}u_R^2,\quad
Z_{\tilde {\varphi }} = 1 -\frac{1}{3\varepsilon}u_R^2,\quad
Z_u = 1 - \frac{3}{4\varepsilon}u_R^2\label{z1}
\end{equation}
from Eqs.~(\ref{zphi}), (\ref{ztphi}), and (\ref{zu}).

\subsection{\label{rmt}Renormalized massless theory}
We have obtained a finite renormalized massless theory by incorporating the dimensional poles to the three $Z$ factors. We can then in principle determine unstable properties using this theory. For example, one can use Eq.~(\ref{g1110}) to find the instability exponent $\eta$ which is just defined as $\Gamma({\bf k},0)\rightarrow |{\bf k}|^{2-\eta}$ for small $|{\bf k}|$ at the instability point as can be seen from Eq.~(\ref{gcs}). Indeed, using Eqs.~(\ref{gar}), (\ref{z}), (\ref{z1}), and (\ref{i1e}), one finds, to one-loop order, Eq.~(\ref{g111e}) becomes
\begin{eqnarray}
\Gamma_{11R}({\bf k},0,\mu)=(Z_{\varphi}Z_{\tilde{\varphi}})^{1/2}\Gamma_{11}({\bf k},0)\nonumber\\
=Z_{\varphi}\lambda_R{\bf k}^2\left\{1+\frac{1}{6\varepsilon}g_3^2N_d|{\bf k}|^{-\varepsilon}\left[1+\frac{7}{12}\varepsilon+O(\varepsilon^2)\right]\right\}\ \nonumber\\
=\lambda_R\left(1+\frac{7}{72}u_R^2\right){\bf k}^2\left\{1-\frac{1}{6}u_R^2\ln\left[\frac{|{\bf k}|}{\mu}\right]+O(u_R^4)\right\}\nonumber\\
\rightarrow\lambda_R\left[1+\frac{7}{72}u_R^2+O(u_R^4)\right]{\bf k}^{2-\eta}\begin{picture}(82,12)\end{picture}\label{g111r}
\end{eqnarray}
with
\begin{equation}
\eta=\frac{1}{6}u_R^2\label{eur}
\end{equation}
for small $\mu$s in equivalence to small momenta if one replaces ${\bf k}$ by $\mu{\bf k}$. In Eq.~(\ref{g111r}), the first line is an example of Eq.~(\ref{gar}) and the second line agrees with Eq.~(\ref{g2r}) for the static vertex function. Accordingly, one sees clearly that the result of the fluctuation-dissipation theorem, Eq.~(\ref{zl1}), is in fact to ensure the correct static limit at $\omega=0$. The third line in Eq.~(\ref{g111r}) exhibits no poles correctly as they just cancel among themselves. We have factored out a non-divergent part here and exponentiated the terms in the braces in the last line. The manipulation appears somehow of brute force to this order but shall be born out by the RG analysis.

Similarly, one finds from Eqs.~(\ref{g11e}) and (\ref{gcs})
\begin{eqnarray}
\Gamma_{11R}({\bf 0},\omega,\mu) &=& i\omega\left\{1-\frac{1}{8}u_R^2\ln\left[\frac{i\omega}{2\lambda_R\mu^2}\right] +O(u_R^3)\right\}\qquad\nonumber\\
&\rightarrow&i\lambda_R\left[1-\frac{1}{8}u_R^2\ln2+O(u_R^3)\right] \left[\frac{\omega}{\lambda_R}\right]^{\frac{2-\eta}{z}}.\label{g1101r}
\end{eqnarray}
As a result, one has
\begin{equation}
z=2-\eta+\frac{1}{4}u_R^2=2+\frac{1}{12}u_R^2\label{zur}
\end{equation}
to one-loop order in the small $\mu$ limit.

These results indicate that one should study whether there is a fixed $u_R^*$ independent of $\mu$ in the small $\mu$ limit. If $u_R^*$ is zero, one recovers the mean-field result. But if there is a finite $u_R^*$ when $\mu\rightarrow0$, one has finite $\eta$ and $z$ which are independent of the scale $\mu$ and are thus universal.

\subsection{\label{rge}RG equations}
In the massless theory one introduces $\mu$ as a substitute for the natural mass scale to define the dimensionless coupling constant $u$. Different values of $\mu$ lead to different renormalized vertex functions. However, they are physically equivalent as they are related to each other by a finite multiplicative renormalization transformation \cite{Justin,amitb,Vasilev}. This can be easily seen from Eq.~(\ref{gar}). The renormalized vertex functions at different $\mu$s are multiplicatively related to the unrenormalized functions of the same given bare theory. So, they are also multiplicatively related by transformation factors that must be finite because the renormalized functions they relate are finite. All such transformations form an RG. The infinitesimal transformations in the case of the massless theory satisfy a differential RG equation. It turns out that this equation leads to useful results such as scaling and universality which lie beyond the perturbation theory.

Now we set up the RG equation. We first consider the case in which $h=0$, i.e., exactly at the instability point. Since the bare functions are independent of $\mu $, one derives the RG equation for the renormalized $\Gamma_{n'nR}$ as
\begin{equation}
\left[ \mu \frac{\partial}{\partial\mu}  + \gamma_{\lambda} \lambda_R
\frac{\partial}{\partial \lambda_R} + \beta \frac{\partial}{\partial u_R}-  \frac{1}{2}n\gamma_{\varphi} -  \frac{1}{2}n'\gamma_{\tilde{\varphi}} \right] \Gamma_{n'nR} = 0
\label{rgenn}
\end{equation}
by differentiating Eq.~(\ref{gar}), where all quantities are renormalized and so
finite and the Wilson functions are defined as derivatives at
constant bare parameters ($\varphi$, $\tilde {\varphi }$, $\lambda$, $g_{3}$)
\begin{eqnarray}
\gamma_{\lambda}(u_R) = \mu \frac{\partial\ln\lambda_R}{\partial\mu}, \qquad \beta (u_R) = \mu \frac{\partial u_R}{\partial\mu}, \nonumber\\
\gamma_{\varphi}(u_R) = \mu \frac{\partial \ln Z_{\varphi}}{\partial\mu}, \qquad\gamma_{\tilde{\varphi}}(u_R)=\mu \frac{\partial \ln Z_{\tilde{\varphi}}}{\partial\mu} .
\label{wil}
\end{eqnarray}
Taking logarithms and then partial derivatives of the equations associated with $\lambda$ and $u$ in Eq.~(\ref{z}) at constant bare parameters and taking into account Eq.~(\ref{zl1}), one finds
\begin{eqnarray}
\gamma_{\lambda}=-\mu \frac{\partial \ln Z_{\lambda}}{\partial\mu} = \frac{1}{2}\gamma_{\tilde{\varphi}}-\frac{1}{2}\gamma_{\varphi},\qquad\label{vmu}\\
\beta(u_R)=-\frac{1}{2}\varepsilon u_R- \mu \frac{\partial \ln Z_{u}}{\partial\mu}u_R\equiv-\frac{1}{2}\varepsilon u_R-\gamma_{u}u_R.\qquad\label{betau}
\end{eqnarray}

Next, we consider the case in which a uniform external field $H_R\neq h_s+rM+gM^3/6$ from Eq.~(\ref{tauh}), i.e., $h_R$ is not exactly at $h_s$. In this case, $m$ is not zero. So, Eq.~(\ref{hs}) becomes
\begin{equation}
\lambda_R h_R(\omega)=\Gamma_{10R}({\bf 0},0;m_R(\omega)).\label{hgm}
\end{equation}
We have included the case in which $m_R$ depends on time. Expanding the right hand side at the instability point at which $m(\tau_s,h_s)=0$ and Eq.~(\ref{hs}) holds, one gets
\begin{equation}
\lambda_R h_R(\omega,\lambda_R,m_R,u_R,\mu)= \sum_{n=1}^{\infty}\frac{1}{n!}\Gamma_{1nR}({\bf 0},0;0)m_R^n.\label{hgm0}
\end{equation}
Therefore, using Eq.~(\ref{rgenn}) and noting that $m=Z_{\varphi}^{1/2}m_R$ similar to $\varphi$, one obtains
\begin{eqnarray}
\left[ \mu \frac{\partial}{\partial\mu}  + \gamma_{\lambda} \lambda_R
\frac{\partial}{\partial \lambda_R} + \beta  \frac{\partial}{\partial u_R}-  \frac{1}{2}\gamma_{\varphi}m_R\frac{\partial}{\partial m_R} - \frac{1}{2} \gamma_{\varphi} \right]h_R \nonumber\\
= 0,\qquad\qquad\qquad\qquad\qquad\qquad\qquad\qquad\qquad\qquad\qquad
\label{rgeh}
\end{eqnarray}
which can also be directly verified by substituting Eq.~(\ref{hgm}) in it.

Another method to derive Eq.~(\ref{rgeh}) is to note that Eq.~(\ref{hgm}) can also be written in the bare form with a bare field $h$. Its transformation can then be found using Eqs.~(\ref{z}) and (\ref{gar}) to be
\begin{equation}
h=Z_{\varphi}^{-1/2}h_R,\label{hr}
\end{equation}
which assures $\lambda h\tilde{\varphi}=\lambda_Rh_R\tilde{\varphi}_R$ and so $\lambda_Rh_R$ is only a shift of the source $\tilde{J}$. Equation~(\ref{hr}) is identical with its static form. Taking this for granted and using Eq.~(\ref{gar}) for $\Gamma_{10}$ again result in Eq.~(\ref{zl1}). Differentiating Eq.~(\ref{hr}) with respect to $\mu$ then leads directly to Eq.~(\ref{rgeh}).

\subsection{\label{sol}Solutions to RG equations, fixed points, and instability exponents}
Equations~(\ref{rgenn}) and (\ref{rgeh}) can be solved by the method of characteristics \cite{Justin,amitb}. We consider first the first equation. Along a characteristic parameterized by $\kappa$ and determined by flow equations and their respective initial conditions
\begin{subequations}
\begin{eqnarray}
\kappa\frac{d\mu(\kappa)}{d\kappa}&=&\mu,\qquad \mu(1)=\mu,\label{kmu}\\
\kappa\frac{d\lambda_R}{d\kappa}&=&\gamma_{\lambda}\lambda_R(\kappa),\qquad\lambda_R(1)=\lambda_R,\label{klr}\\
\kappa\frac{du_R(\kappa)}{d\kappa}&=&\beta[u_R(\kappa)],\qquad u_R(1)=u_R,\label{kur}
\end{eqnarray}
\end{subequations}
$\Gamma_{n'n}$ satisfies
\begin{equation}
\kappa\frac{d\Gamma_{n'nR}}{d\kappa}=\left(\frac{1}{2}n\gamma_{\varphi} +  \frac{1}{2}n'\gamma_{\tilde{\varphi}}\right)\Gamma_{n'nR},
\end{equation}
whose solution is
\begin{widetext}
\begin{equation}
\Gamma_{n'nR}(\{{\bf k}_i\},\{\omega_i\};\lambda_R(\kappa),u_R(\kappa),\mu\kappa)= \Gamma_{n'nR}(\{{\bf k}_i\},\{\omega_i\};\lambda_R,u_R,\mu) \exp\left\{\frac{1}{2}\int_1^{\kappa}\frac{dx}{x}\left[n\gamma_{\varphi}(x) +  n'\gamma_{\tilde{\varphi}}(x)\right]\right\},\label{gnnc}
\end{equation}
where we have used the solution of Eq.~(\ref{kmu}), $\mu(\kappa)=\mu\kappa$.

On the other hand, from the na\"{\i}ve dimensional analysis in Sec.~\ref{dimensions}, one has a generalized homogenous relation
\begin{equation}
\Gamma_{n'nR}(\{{\bf k}_i\},\{\omega_i\};\lambda_R,u_R,\mu)= \kappa^{d_{n'n}}\lambda_R\Gamma_{n'nR}(\{{\bf k}_i/\kappa\},\{\omega_i/\lambda_R\kappa^2\};u_R,\mu/\kappa).\label{gnnh}
\end{equation}
Note the $\lambda_R$ factor from Table~\ref{dim} for dimensional reasons. Applying this equation to the left hand side of Eq.~(\ref{gnnc}), one finds
\begin{equation}
\Gamma_{n'nR}(\{\kappa{\bf k}_i\},\{\omega_i\};\lambda_R,u_R,\mu)= \lambda_R(\kappa)\Gamma_{n'nR}(\{{\bf k}_i\},\{\omega_i/\lambda_R(\kappa)\kappa^2\};u_R(\kappa),\mu)\kappa^{d_{n'n}} \exp\left\{-\frac{1}{2}\int_1^{\kappa}\frac{dx}{x}\left[n\gamma_{\varphi} +  n'\gamma_{\tilde{\varphi}}\right]\right\},\label{gnnk}
\end{equation}
where we have relabeled ${\bf k}$ with $\kappa{\bf k}$. The $\mu$ dependence here is totally free and can be suppressed. In fact, one can simply set $\kappa=\mu$ and $\mu$ itself, the value at $\kappa=1$, just 1 at the beginning.

Equation~(\ref{gnnk}) relates vertex functions at different wavenumbers. In particular, it quantifies how the vertex functions change when the wavenumbers are reduced. No simple scaling occurs, however, as the momenta are rescaled. The reverse happens when the coupling constant reaches a value such that any further rescaling does not affect it. This value is found at the fixed points satisfying
\begin{equation}
\beta(u_R^*)=0.\label{bu0}
\end{equation}
At such a fixed point, $u_R$ does not change with $\kappa$ from Eq.~(\ref{kur}). As they depend on $\kappa$ through $u_R$, $\gamma_{\varphi}$, $\gamma_{\tilde{\varphi}}$, and $\gamma_{\lambda}$ also assume their fixed point values, which will all be marked by stars. As a result, Eq.~(\ref{klr}) leads to
\begin{equation}
\lambda_R(\kappa)=\lambda_R\kappa^{\gamma_{\lambda}^*}\label{lrk}
\end{equation}
and, using Eq.~(\ref{vmu}) for $\gamma_{\lambda}$, Eq.~(\ref{gnnk}) becomes
\begin{equation}
\Gamma_{n'nR}(\{\kappa{\bf k}_i\},\{\omega_i\};\lambda_R,u_R^*)= \lambda_R\Gamma_{n'nR}(\{{\bf k}_i\},\{\omega_i/\lambda_R\}\kappa^{-2-\gamma_{\lambda}^*};u_R^*)\kappa^{d_{n'n}- \frac{1}{2}(n+1)\gamma_{\varphi}^* - \frac{1}{2} (n'-1)\gamma_{\tilde{\varphi}}^*},\label{gnnks}
\end{equation}
which exhibits exact scaling. In particular,
\begin{equation}
\Gamma_{11R}(\kappa{\bf k},\omega;\lambda_R,u_R^*)= \lambda_R\Gamma_{11R}({\bf k},(\omega/\lambda_R)\kappa^{-2-\gamma_{\lambda}^*};u_R^*)\kappa^{2- \gamma_{\varphi}^*}\label{g2ks}
\end{equation}
\end{widetext}
using Table~\ref{dim} for $d_{n'n}$. Comparing with Eq.~(\ref{gcs}), one finds that
\begin{subequations}
\label{exponent}
\begin{eqnarray}
z=2+\gamma_{\lambda}^*=2+\frac{1}{2} \gamma_{\tilde{\varphi}}^*-\frac{1}{2}\gamma_{\varphi}^*,\label{zs}\\
\eta=\gamma_{\varphi}^*,\label{etas}\\
\gamma_{\tilde{\varphi}}^*=\eta+2z-4\label{gtp}
\end{eqnarray}
\end{subequations}
using Eq.~(\ref{vmu}).
Therefore, Eq.~(\ref{g2ks}) becomes
\begin{equation}
\Gamma_{11R}(\kappa{\bf k},\omega;\lambda_R,u_R^*)= \lambda_R\Gamma_{11R}({\bf k},(\omega/\lambda_R)\kappa^{-z};u_R^*)\kappa^{2- \eta}.\label{g2ksf}
\end{equation}
For $\omega=0$, ${\bf k}={\bf 1}$, and $\kappa=|{\bf k}|$, Eq.~(\ref{g2ksf}) reads
\begin{equation}
\Gamma_{11R}({\bf k},0;\lambda_R,u_R^*)= \lambda_R\Gamma_{11R}({\bf 1},0;u_R^*)|{\bf k}|^{2- \eta},
\end{equation}
which is in fact Eq.~(\ref{g111r}). For ${\bf k}={\bf 0}$ and $\kappa=(\omega/\lambda_R)^{1/z}$, it becomes
\begin{equation}
\Gamma_{11R}({\bf 0},\omega;\lambda_R,u_R^*)= \lambda_R\Gamma_{11R}({\bf 0},1;u_R^*)(\omega/\lambda_R)^{(2- \eta)/z},
\end{equation}
which can be shown to be just Eq.~(\ref{g1101r}). Accordingly, one has
\begin{eqnarray}
\Gamma_{11R}({\bf k},0;\lambda_R,u_R^*)&=&
\lambda_R\left(1-\frac{7}{108}\varepsilon\right)|{\bf
k}|^{2+\frac{\varepsilon}{9}},\qquad\\
\Gamma_{11R}({\bf 0},\omega;\lambda_R,u_R^*)&=&
i\lambda_R\left(1+\frac{1}{12}\varepsilon\right)
\left(\frac{\omega}{\lambda_R}\right)^{1+\frac{\varepsilon}{12}}\qquad
\end{eqnarray}
to one loop using Eqs.~(\ref{ufp}) and (\ref{exp1}) below.

Similarly, defining one more flow equation associated with $m_R$ as
\begin{equation}
\kappa\frac{dm_R(\kappa)}{d\kappa}=-\frac{1}{2}\gamma_{\varphi}m_R,\qquad m_R(1)=m_R,\label{kmr}
\end{equation}
one arrives at the counterpart of Eq.~(\ref{gnnk}) as
\begin{eqnarray}
h_R(t,\lambda_R,m_R,u_R)=\kappa^{(d+2)/2}
\exp\left\{-\frac{1}{2}\int_1^{\kappa}\frac{dx}{x}\gamma_{\varphi}
\right\}\nonumber\\
\times
h_R(\lambda_R(\kappa)t\kappa^{2},m_R(\kappa\kappa^{-(d-2)/2},u_R(\kappa)),\quad
\end{eqnarray}
where we have switched to the time domain and used
Table~\ref{dim}. At the fixed point,
\begin{equation}
m_R(\kappa)=m_R\kappa^{-\frac{1}{2}\eta}\label{mrk}
\end{equation}
from Eq.~(\ref{kmr}) and hence
\begin{equation}
h_R(t,\lambda_R,m_R,u_R^*)=\kappa^{\beta\delta/\nu}
h_R(\lambda_Rt\kappa^{z},m_R\kappa^{-\beta/\nu},u_R^*)\label{hrs}
\end{equation}
with
\begin{eqnarray}
\beta\delta/\nu&=&\frac{1}{2}(d+2-\eta),\label{bdn}\\
\beta/\nu&=&\frac{1}{2}(d-2+\eta)\label{bn}\\
\delta&=&\frac{d+2-\eta}{d-2+\eta},\label{des}
\end{eqnarray} using
Eqs.~(\ref{lrk}) and (\ref{exponent}). Choosing $\kappa\sim
m_R^{\nu/\beta}$ in Eq.~(\ref{hrs}) leads to a scaling form for
the equation of state at the instability point
\begin{equation}
h_R(t,\lambda_R,m_R)=m_R^{\delta}f_3(\lambda_Rtm_R^{\nu
z/\beta})\label{hmt}
\end{equation}
where $f_3$ is a scaling function.

We have solved the RG equations and studied the behavior of a system whose renormalized coupling constant lies exactly at the fixed point. We now investigate how and when the renormalized coupling constant flows into the fixed point.

Expanding $\beta(u_R)$ in the vicinity of a simple fixed point $u_R^*$,
\begin{equation}
\beta(u_R)=\beta'(u_R-u_R^*),\label{brs}
\end{equation}
one finds from Eq.~(\ref{kur})
\begin{equation}
u_R(\kappa)-u_R^*=(u_R-u_R^*)\kappa^{\beta'},\label{uks}
\end{equation}
where $\beta'$ is the derivative of $\beta$ at $u_R^*$. One sees therefore that if $\beta'>0$, $u_R(\kappa)$ will flow to $u_R^*$ for $\kappa\rightarrow0$ independent on the initial renormalized coupling constant. On the other hand, if $\beta'<0$, $u_R(\kappa)$ will flow to $u_R^*$ for $\kappa\rightarrow\infty$ for arbitrary initial renormalized coupling constant near the fixed point. One calls the fixed point in the first case infrared stable and that in the second case ultraviolet stable. The reason is that one sees from Eq.~(\ref{gnnks}) that, as $\kappa\rightarrow0$, one probes smaller and smaller momenta and thus the large scale behavior.

We have computed all necessary $Z$s in Eq.~(\ref{z1}) to one-loop order. Accordingly, Eqs.~(\ref{wil}) and (\ref{betau}) result in
\begin{eqnarray}
\beta(u_R)=-\frac{1}{2}\varepsilon u_R- \frac{3}{4}u_R^{3},\label{betau1}\\
\gamma_{\varphi}=\beta(u_R)\frac{\partial\ln Z_{\varphi}}{\partial u_R}=\frac{1}{6}u_R^2,\label{gpu1}\\
\gamma_{{\tilde{\varphi}}}=\beta(u_R)\frac{\partial\ln Z_{{\tilde{\varphi}}}}{\partial u_R}=\frac{1}{3}u_R^2\label{gtpu1}.
\end{eqnarray}
Eq.~(\ref{betau1}) has three fixed points. The Gaussian fixed point $u_R^* = 0$ is infrared stable for $\varepsilon<0$ or $d>6$ and ultraviolet stable for $\varepsilon>0$. The other two purely imaginary conjugate fixed points are
\begin{equation}
u_R^{*2} = -\frac{2}{3}\varepsilon. \label{ufp}
\end{equation}
Although they are imaginary in values, they are infrared stable and thus control the large scale behavior for $\varepsilon>0$ or $d<6$. Moreover, at these imaginary fixed points,
\begin{equation}
\eta=-\frac{1}{9}\varepsilon, \qquad z=2-\frac{1}{18}\varepsilon, \qquad \delta=2+\frac{1}{3}\varepsilon \label{exp1}
\end{equation}
from Eqs.~(\ref{gpu1}), (\ref{gtpu1}), (\ref{exponent}), and
(\ref{des}), all are real. In addition, the first two exponents
agree with the results found in Eqs.~(\ref{eur}) and (\ref{zur})
provided the fixed point value in Eq.~(\ref{ufp}) is substituted.
For $\varepsilon=0$ or $d = 6$ corresponding to $u_R^*=0$, all the
exponents recover their Gaussian fixed point values listed in
Table~\ref{mfexp}. This fact shows again that the fixed points,
though imaginary, describe correctly the large-scale fluctuations
at the instability point at which the real mean-field transition
takes place. Moreover, using a finite-time scaling with Monte
Carlo RG method, we have found that the imaginary fixed point can
affect the RG flows in the temperature-driven FOPTs of the Potts
models in $d=2$ \cite{Fan11}. From these exponents, others can
also be obtained. We shall return to the instability exponents in
Sec.~\ref{YL} below.

\subsection{\label{fts}Finite-time scaling}
Although the scaling and universality behavior found in the last section is similar to the critical phenomena, it may not be easily observable because of the instability and the dynamic nature of the transition. An accessible method is to employ finite-time scaling which has been proved to be effective in the critical phenomena \cite{gong,zhongintech}. In this method, one varies the external field linearly through the transition point to probe the scaling behavior. We derive the finite-time scaling form for the FOPTs in this section.

Let the rate $R$ of the linear driving transform as
\begin{equation}
R=Z_rR_R,\label{zr}
\end{equation}
one finds
\begin{equation}
Z_r=Z_{\lambda}^{-1}Z_{\varphi}^{-1/2}=Z_{\varphi}^{-1}Z_{\tilde{\varphi}}^{1/2}\label{zrlp}
\end{equation}
by assuming the driving form, Eq.~(\ref{hRt}) with the $\lambda$ included, is identical in both the bare and the renormalized forms such that
\begin{equation}
h_R=\lambda_RR_Rt=Z_{\varphi}^{1/2}h=Z_{\varphi}^{1/2}\lambda Rt
\end{equation}
using Eq.~(\ref{hr}) and Eqs.~(\ref{z}) and (\ref{zl1}). Note that although such a driving breaks the time translational symmetry, it has been shown that this only introduces a possible initial slip near criticality arising from nonequilibrium initial conditions that induce a new singularity \cite{Janssen89}. In finite-time scaling, we always start the driving far away from the critical/instability point as mentioned in Sec.~\ref{mfhys}. Consequently, the initial slip should show no effects except that one starts driving near the point to study it purposely.  Without considering the initial slip, the driving brings on no new singularity \cite{zhong06}. Accordingly, the three $Z$ factors introduced in Sec.~\ref{rgsrc} suffice for curing all the intrinsic singularities. This is why $Z_r$ is related to the previous ones as seen in Eq.~(\ref{zrlp}). As $\lambda_R$ and $R_R$ are related, we choose $R_R$ as a variable and write the RG equation for $h_R$ as
\begin{eqnarray}
\left[ \mu \frac{\partial}{\partial\mu}  + \gamma_{r} R_R
\frac{\partial}{\partial R_R} + \beta  \frac{\partial}{\partial
u_R}-  \frac{1}{2}\gamma_{\varphi}m_R\frac{\partial}{\partial m_R}
- \frac{1}{2} \gamma_{\varphi} \right]h_R\nonumber\\
 = 0\qquad\qquad\qquad\qquad\qquad\qquad\qquad\qquad\qquad\qquad\qquad,
\label{rgehr}
\end{eqnarray}
with
\begin{equation}
\gamma_{r}=\mu\frac{\partial\ln R_R}{\partial\mu}=-\mu\frac{\partial\ln Z_r}{\partial\mu}=\gamma_{\varphi}-\frac{1}{2}\gamma_{\tilde{\varphi}}\label{gr}
\end{equation}
from Eqs.~(\ref{zr}), (\ref{zrlp}), and (\ref{wil}). The solution at the fixed point is then
\begin{equation}
h_R(R_R,m_R,u_R^*)= \kappa^{\beta\delta/\nu}
h_R(R_R\kappa^{-r_H},m_R\kappa^{-\beta/\nu},u_R^*)\label{hrsr}
\end{equation}
with
\begin{equation}
r_H=\frac{d+6}{2}-\gamma_{r}^*=z+\frac{1}{2}(d+2-\eta)=z+\beta\delta/\nu\label{rh}
\end{equation}
by noting that
\begin{equation}
\kappa\frac{dR_R(\kappa)}{d\kappa}=\gamma_{r}R_R(\kappa), \qquad R_R(1)=R_R, \label{krr}
\end{equation}
and $[R]=[h]-[\lambda t]=(d+6)/2$ using Table~\ref{dim}. Uses have
also been made of Eqs.~(\ref{exponent}) and (\ref{bdn}) and the
solution of Eq.~(\ref{krr}) at the fixed point.
Equation~(\ref{rh}) is just Eq.~(\ref{sl}) and Eq.~(\ref{hrsr}) is
just another form of Eq.~(\ref{mscaling}). In fact, one can derive
directly an RG equation for $m_R$ with $h_R$ as its variable
\cite{zhongl05}. As a result, one finds
\begin{eqnarray}
n_H=\beta\delta/\nu r_H=\frac{d+2-\eta}{d+2-\eta+2z},\nonumber\\
n_m=\beta/\nu r_H=\frac{d-2+\eta}{d+2-\eta+2z},\label{nhm}
\end{eqnarray}
whose one-loop $\varepsilon$ expansions are
\begin{equation}
n_H=\frac{2}{3}-\frac{1}{54}\varepsilon,\qquad n_m=\frac{1}{3}-\frac{7}{108}\varepsilon,
\end{equation}
respectively, using Eq.~(\ref{exp1}), which recover their mean-field values for $\varepsilon=0$. These hysteresis exponents were found to be comparable with direct numerical solutions of Eqs.~(\ref{H}) and (\ref{de}) \cite{zhongl05}. This again supports the relevance of the $\varphi^3$ theory to the FOPTs. We shall return to this comparison in Sec.~\ref{YL} below.

\subsection{\label{cts}Leading corrections to scaling}
When the initial $u_R$ does not lie at $u_R^*$ but is near it, there are leading corrections to the exact scaling behavior as exhibited in Eqs.~(\ref{gnnks}) and (\ref{hrs}).

Consider the external field for an instance and assume that $\beta$ is given by Eq.~(\ref{brs}). Let
\begin{eqnarray}
\ln Z_{\varphi}(u_R)=-\int_{u_R^*}^{u_R}\frac{dx}{\beta(x)}[\gamma_{\varphi}(x)-\eta], \qquad\label{zurc}\\
m_R(u_R)=m_RZ_{\varphi}^{-1/2}(u_R),\qquad\label{murc}\\
\lambda_R(u_R)=\lambda_R\exp\left\{-\int_{u_R^*}^{u_R}\frac{dx}{\beta(x)} [\gamma_{\lambda}(x)-z+2]\right\},\qquad\\
\tilde{u}=(u_R-u_R^*)\exp\left\{\int_{u_R^*}^{u_R}dx\left[\frac{\beta'}{\beta(x)}- \frac{1}{x-u_R^*}\right]\right\},\qquad\label{zrc}
\end{eqnarray}
which are finite renormalizations that eliminate trivial deviations from the fixed point theory and correspond simply to a change of normalization of the different scaling variables \cite{Justin}. Then, we assume
\begin{widetext}
\begin{equation}
h_R(t,\lambda_R,m_R,u_R,\mu)=Z_{\varphi}^{-1/2}(u_R)h_R(t,\lambda_R(u_R),m_R(u_R),u_R^*,\mu) U(t,\lambda_R(u_R),m_R(u_R),\tilde{u},\mu)
\end{equation}
with the boundary condition $U(t,\lambda_R(u_R),m_R(u_R),0,\mu)=1$ and substitute it into Eq.~(\ref{rgeh}), resulting in
\begin{eqnarray}
\left[ \mu \frac{\partial}{\partial\mu}  + (z-2) \lambda_R
\frac{\partial}{\partial \lambda_R} + \beta'\tilde{u} \frac{\partial}{\partial \tilde{u}}-  \frac{1}{2}\eta m_R\frac{\partial}{\partial m_R} \right]U(t,\lambda_R(u_R),m_R(u_R),\tilde{ u},\mu) = 0,\\
\left[ \mu \frac{\partial}{\partial\mu}  + (z-2) \lambda_R
\frac{\partial}{\partial \lambda_R} -  \frac{1}{2}\eta m_R\frac{\partial}{\partial m_R}-\frac{1}{2}\eta \right]h_R(t,\lambda_R(u_R),m_R(u_R),u_R^*,\mu) = 0.
\label{rgec}
\end{eqnarray}
These equations can be solved again by the method of characteristics after expanding $U$ in powers of $\tilde{u}$ with the result
\begin{equation}
h_R(t,\lambda_R,m_R,u_R)=Z_{\varphi}^{-1/2}(u_R)\kappa^{\beta\delta/\nu} h_R(\lambda_Rt\kappa^{z},m_R\kappa^{-\beta/\nu},u_R^*)\left[1+\sum_{s=1}^{\infty}\tilde{u}^s \kappa^{s\beta'}f_4(\lambda_Rt\kappa^{z},m_R\kappa^{-\beta/\nu})\right],\label{hcor}
\end{equation}
where $f_4$ is a scaling function and uses have been made of Eqs.~(\ref{bdn}) and (\ref{bn}). A finite-time scaling form of Eq.~(\ref{hcor}) is
\begin{equation}
h_R(R,\lambda_R,m_R,u_R)=Z_{\varphi}^{-1/2}(u_R)R^{\beta\delta/\nu r_H} h_R(1,m_RR^{-\beta/\nu r_H},u_R^*)\left[1+\sum_{s=1}^{\infty}\tilde{u}^s R^{s\beta'/r_H}f_5(m_RR^{-\beta/\nu r_H})\right]\label{hcorr}
\end{equation}
\end{widetext}
with another scaling function $f_5$.

Note that the exponent of the corrections, $\beta'$, depends only on the fixed point and is thus universal. To one-loop order,
\begin{equation}
\beta'=\varepsilon
\end{equation}
from Eqs.~(\ref{betau1}) and (\ref{ufp}). It is real. However, the derivative of $\gamma_{\varphi}$ with respect to its argument at $u_R^*$,
\begin{equation}
\gamma_{\varphi}'=\frac{1}{3}u_R^*=\pm i\frac{1}{3}\sqrt{\frac{2}{3}\varepsilon}
\label{gp}
\end{equation}
from Eqs.~(\ref{gpu1}) and (\ref{ufp}), is imaginary. Expanding $\beta$ and $\gamma_{\varphi}$ to first order in $u_R-u_R^*$, one finds from Eq.~(\ref{zurc})
\begin{equation}
\ln Z_{\varphi}(u_R)=-\frac{\gamma_{\varphi}'}{\beta'}(u_R-u_R^*)= \mp i\sqrt{\frac{2}{27\varepsilon}}(u_R-u_R^*),
\end{equation}
which may be complex depending on $u_R$.

\section{\label{YL}Existent $\varphi^3$ theories, Yang--Lee edge singularity, and resummed instability exponents}
We have studied in detailed the field-theoretical RG theory of the $\varphi^3$ theory of the FOPTs in the last section. In fact, Hamiltonians that possess such a cubic interaction have been utilized to model many phase transitions.
Examples include the isotropic to nematic phase transition in liquid crystals \cite{Gennes,Priest}, systems described by the Potts model \cite{Potts,Zia,Priest,Amit76}, in particular the percolation problem \cite{Harris75,Amit76,Priest} which is the single state Potts model \cite{Fortuin}, the Edwards-Anderson model of spin glasses \cite{Edwards,Harris76}, a lot of nonequilibrium systems \cite{Odor} such as the dynamic isotropic percolation and the directed percolation \cite{Janssen05}, the Reggeon field theory for high-energy scattering amplitudes \cite{Abarbanel,Moshe} which falls in the same universality class as the directed percolation \cite{Cardy80}, the Yang--Lee edge singularity \cite{Fisher78} and its related problems such as isotropic branched polymers in a good solvent and undirected lattice animals, Anderson localization, and directed branched polymers and directed lattice animals, which relate to the edge singularity in $d+2$ dimensions \cite{Parisi81}, $d+2$ dimensions \cite{Lubensky81}, and $d+1$ dimensions \cite{Cardy82}, respectively, as well as quantum field theory models in particle physics \cite{McKane76}.
For the case of $\varphi^3$ models that are directly related to the one studied here, they have been classified into two categories depending on whether `unphysical' limits such as a particular limit of state numbers and a purely imaginary coupling constant have to be taken or not \cite{Alcantara}. Field theories in the category with real Hamiltonians and without a particular state number limit have an unstable ground state and instanton solutions such that a perturbation expansion is not an adequate approach \cite{McKane79}. Moreover, the $\phi^4$ interaction becomes relevant as the space dimensionality lowers from six \cite{McKane79}. These are not true for theories in the second category. Thus, the perturbation series has oscillatory terms and hence is Borel summable \cite{Houghton,kirkham} and the quartic interaction has been found to be irrelevant \cite{Amit77,Elderfield,kirkham}. The Yang--Lee edge singularity belongs to this class and thus the $\varepsilon$ expansion in the RG analysis provides good results \cite{Alcantara}. In the following, we shall show that the dynamics of FOPTs near their instability points falls in the same universality class to the Yang--Lee edge singularity and shall apply its existent results to extend ours for the FOPTs.

\subsection{\label{YLes}Yang--Lee edge singularity}
According to Yang and Lee \cite{Yang}, to study the problem of phase transitions, it is necessary to study the distribution in the complex fugacity plane of the roots of the grand partition function. Under a class of general conditions, these roots lie on a circle in the plane for the Ising model and its equivalent lattice gas model \cite{LeeY}. For the Ising model, the fugacity is proportional to $\exp(-2H)$ (note that our definition of $H$ has absorbed in it the thermal factor) and so the Lee--Yang circle theorem places the zeros at imaginary magnetic fields. For $T<T_c$, there are zeroes at $H=0$ in the thermodynamic limit and the magnetization as a function of $H$ exhibits a jump at $H=0$; while for $T>T_c$, there is a gap of width $2iH_0(T)$ within which no zeros exist at all. It was found that the distribution of the zeros was singular at the edge of the gap \cite{Kortman}. This Yang--Lee edge singularity is described by \cite{Fisher78}
\begin{equation}
m=M-M_0\sim(H-iH_0)^\sigma \label{yles}
\end{equation}
with
\begin{equation}
\sigma=\frac{1}{\delta}=\frac{d-2+\eta}{d+2-\eta},\label{sigma}
\end{equation}
where $M_0$ is the magnetization at the imaginary field $iH_0$.

Consider a continuous spin Ising model in an imaginary external field above its critical temperature, viz., Eq.~(\ref{H}) with an imaginary $H$ and $r>0$ \cite{Fisher78,Cardy82}. As there is no spontaneous symmetry breaking for $r>0$ but only the imaginary external field, the shift in Eq.~(\ref{pmvp}) can be only to an imaginary $M$ induced by $H$. One sees then from Eqs.~(\ref{tauh}) and (\ref{g3}) that $\tau$ is real but both $h$ and $g_3$ are imaginary. Therefore, the leading infrared behavior of the Yang--Lee edge singularity is governed by Eq.~(\ref{hp}) with purely imaginary $g_3$ and $h$ \cite{Fisher78}. A redefinition of $\varphi$ to $i\varphi$ \cite{Parisi81} which is a dummy variable that will be integrated out then turns the Hamiltonian back to exactly the form of Eq.~(\ref{hp}). Therefore, this singularity for $T>T_c$ is indeed described by the same model as the FOPTs below $T_c$. As a consequence, for $d>d_c=6$, the classical mean-field theory results in $\sigma=1/2$ from Eq.~(\ref{sigma}) and Table~\ref{mfexp}. For $d<6$, the $\varepsilon=6-d$ expansions of the RG functions for the Yang--Lee edge singularity have been computed up to three loops from those of the Potts model. Their resummed exponents agree impressively with those of the high-temperature series analysis \cite{Kurtze} even down to the exact result of $\sigma=-1/2$ for the one dimensional Ising model where $\varepsilon=5$ \cite{Alcantara}. Also the dynamic critical exponent for the Yang--Lee edge singularity has been computed up to two-loop order \cite{Breuer81}.

From the fixed point given by Eq.~(\ref{ufp}), one sees that fixed point value of $u_R$ itself is thus imaginary. This again shows that the infrared behavior of the FOPT at its instability fixed point is just described by the same field theory as that of the Yang--Lee edge singularity albeit in opposite temperature ranges. We can then check that all our RG functions agree with previous results \cite{Amit76,Alcantara,Breuer81}. Moreover, we now employ these results to estimate the instability exponents in the following section.

\subsection{Resummed instability exponents}
\begin{table*}
\caption{\label{exp} Instability exponents.}
\begin{ruledtabular}
\begin{tabular}{llllllll}
$d$   &6\footnotemark[1]&\qquad5       &\qquad4       &\qquad3       &\qquad2       &\qquad1   &\quad0\\
\hline
$\eta$\footnotemark[2]&0&$-0.157$&$-0.351$&$-0.561$&$-0.778$&$-1$\footnotemark[3]&$-1.224$\\
$\eta$\footnotemark[4]&0&$-0.147\pm0.002$&$-0.329^{+0.012}_{-0.013}$&$-0.527^{+0.029}_{-0.033}$ &$-0.736^{+0.053}_{-0.061}$&$-0.952^{+0.083}_{-0.098}$&\\
$\eta$\footnotemark[5] &0&$-0.147\pm0.002$&$-0.329^{+0.012}_{-0.013}$&$-0.527^{+0.029}_{-0.033}$ &$-0.747^{+0.064}_{-0.050}$ &$-1$\footnotemark[3]&$-1.224$\\
$\delta$\footnotemark[6]&2&\quad\!$2.505\pm0.003$&\quad\!$3.788^{-0.034}_{+0.038}$ &~$11.685^{-0.733}_{+0.951}$&$-6.355^{-0.502}_{+0.336}$&$-2$&$-1$\footnotemark[7]\\
$\nu$\footnotemark[6]&1/2&\quad\!$0.701\pm0.0005$&\quad\!$1.197^{-0.009}_{+0.009}$ &\quad\!$4.228^{-0.244}_{+0.317}$&$-2.677^{-0.251}_{+0.168}$&$-1$&$-0.620$\\
$\nu/\beta\delta$\footnotemark[6]&1/4&\quad\!$0.280\pm0.0001$&\quad\!$0.316\pm0.0006$ &\quad\!$0.362\pm0.002$& \quad\!$0.421^{+0.006}_{-0.004}$&\quad\!$0.5$&\quad\!$0.620$\\
$\gamma/\beta\delta$\footnotemark[6]&1/2&\quad\!$0.580\pm0.0001$&\quad\!$0.684\pm0.0006$ &\quad\!$0.819\pm0.001$& \quad\!$1$&\quad\!$1.25$&\quad\!$1.620$\\
$z$\footnotemark[8]&2&$\quad\!1.938$&$\quad\!1.874$&$\quad\!1.809$&$\quad\!1.743$&$\quad\!1.677$ &$\quad\!1.612$\\
$z$\footnotemark[9]&2&$\quad\!1.944$&$\quad\!1.885$&$\quad\!1.825$&$\quad\!1.763$&$\quad\!1.699$ &$\quad\!1.633$\\
$z$\footnotemark[5]&2&$\quad\!1.941\pm0.003$&$\quad\!1.880\pm0.006$&$\quad\!1.817\pm0.008$& $\quad\!1.753\pm0.010$&$\quad\!1.677$ &$\quad\!1.612$\\
$r_H$\footnotemark[6]&6&\quad\!$5.512\pm0.003$&\quad\!$5.038^{-0.006}_{+0.007}$& \quad\!$4.572^{-0.015}_{+0.017}$ &\quad\!$4.117^{-0.032}_{+0.025}$&\quad\!3.678&\quad\!3.224\\
$n_H$\footnotemark[6]&$2/3$&\quad\!$0.648\pm0.0004$&\quad\!$0.627\pm0.001$ &\quad\!$0.603\pm0.003$ &\quad\!$0.575^{+0.004}_{-0.005}$&\quad\!0.544&\quad\!0.5\\
$n_m$\footnotemark[6]&$1/3$&\quad\!$0.259^{+0.0002}_{-0.0003}$&\quad\!$0.166^{+0.001}_{-0.002}$ &\quad\!$0.0516^{+0.0033}_{-0.0039}$ &$-0.0905^{+0.0073}_{-0.0057}$&$-0.272$&$-0.5$\\
$n_H$\footnotemark[10]&0.654(8)&\quad\!$0.645(3)$&\quad\!$0.625(12)$ &\quad\!$0.595(30)$ &&\\
$n_m$\footnotemark[10]&0.34(6)&\quad\!$0.27(4)$&\quad\!$0.17(8)$ &&&\\
\end{tabular}
\end{ruledtabular}
\footnotetext[1]{Mean-field results.}
\footnotetext[2]{Equation~(\ref{e3s}) using the [3/2] Pad\'{e} approximant.}
\footnotetext[3]{Exact result \cite{Fisher78}.}
\footnotetext[4]{Reference~\cite{Alcantara} (quoted errors reflect the spread in different resummations).}
\footnotetext[5]{Final estimates (quoted errors reflect the spread in different resummations).}
\footnotetext[6]{Quoted errors reflect the corresponding spreads in $\eta$ and/or $z$ in the final estimates.}
\footnotetext[7]{Exact result \cite{Breuer81,Parisi81}.}
\footnotetext[8]{Equation~(\ref{z2s}) using the [2/1] Pad\'{e} approximant.}
\footnotetext[9]{Equation~(\ref{z11s}) using the [1/1] Pad\'{e} approximant.}
\footnotetext[10]{Numerical results \cite{zhongl05}.}
\end{table*}
The three- and two-loop results for $\eta$ and $z$ of the Yang--Lee edge singularity are \cite{Alcantara,Breuer81}
\begin{eqnarray}
\eta=-\frac{1}{9}\varepsilon-\frac{43}{3^6}\varepsilon^2+ \left(\frac{16\zeta(3)}{3^5}-\frac{8375}{2^23^{10}}\right)\varepsilon^3+O(\varepsilon^4),\qquad\label{e3}\\
z=2-\frac{1}{18}\varepsilon+\left(\frac{241}{11664}-\frac{1}{8}\ln\frac{4}{3}\right)\varepsilon^2 +O(\varepsilon^3),\qquad\label{z2}
\end{eqnarray}
respectively, where $\zeta$ is the Riemann function. The former series has been resummed using a [2/1] Pad\'{e} approximant, a [2/1] Pad\'{e}--Borel method, and a conformal mapping technique to provide rather good estimates for even $d=1$ or $\varepsilon=5$ \cite{Alcantara}. The latter has also been resummed by a [2/1] Pad\'{e} approximant with the aid of the result of $z$ in $d=0$ given by
\begin{equation}
z(d=0)=2-\frac{1}{2}[2+\eta(d=0)],\label{zd0}
\end{equation}
in which $\eta(d=0)$ is obtained by a similar approximation to two-loop order \cite{Breuer81}. Such a [2/1] Pad\'{e} approximant to $\eta$ has been performed using the exact result of $\eta=-1$ in $d=1$ for the Yang--Lee edge singularity to two-loop order \cite{Kurtze}.

As we now have a three-loop order of $\eta$, we can form a [3/2] Pad\'{e} approximant to $\eta$ using the same exact result in $d=1$. The result is
\begin{equation}
\eta=-\frac{\varepsilon}{9}\frac{1-0.908\varepsilon-2.223\varepsilon^2} {1-1.144\varepsilon-1.066\varepsilon^2}.\label{e3s}
\end{equation}
We list in Table~\ref{exp} the values for various $d$s together with the averages of those from various other resummation methods \cite{Alcantara}. One sees that the former is a bit larger than the latter for large $d$s. This is expected as the value in $d=1$ has been fixed. Accordingly, as our final estimates, the results from Ref.~\cite{Alcantara} are kept for $d\geq3$ and are averaged with the present one in $d=2$, giving rise to a closer value to $\eta(d=2)=-0.78(2)$ from the high-temperature series expansion \cite{Kurtze}. From these values, we can then compute $\delta$ using Eq.~(\ref{des}). In addition, we have also collected in Table~\ref{exp} the values of $\nu$ from
\begin{equation}
\nu=\frac{2}{d-2+\eta}
\end{equation}
from Eq.~(\ref{bn}), since it has been shown that $\beta=1$ for the scalar $\varphi^3$ model \cite{Alcantara,Reeve}. For the same reason, from
\begin{equation}
\gamma=\beta(\delta-1),\label{gbd}
\end{equation}
$\gamma$ is just given by $\delta-1$. However, the values of $\nu$ and $\gamma$ and even $\delta$ look quite strange for lower dimensions. In fact, as we are studying the field-driven case, we would consider the field instead of the temperature deviations away from the instability point. The corresponding relations are then
\begin{equation}
\xi\sim h_R^{-\nu/\beta\delta},\qquad\chi\sim h_R^{-\gamma/\beta\delta}
\end{equation}
with
\begin{equation}
\nu/\beta\delta=\frac{2}{d+2-\eta}\qquad
\gamma/\beta\delta=1-\sigma
\end{equation}
from Eqs.~(\ref{bdn}) and (\ref{sigma}) and (\ref{gbd}). These two exponent ratios, which correspond simply to $\nu$ and $\gamma$, respectively, in the context of branched polymers \cite{Parisi81} and directed animals \cite{Cardy82}, appear normal as seen in Table~\ref{exp}.

Equation~(\ref{e3s}) gives a slightly different $\eta$ and hence $z$ in $d=0$ as compared to $z=1.614$ from the two-loop result \cite{Breuer81}. As a consequence, the [2/1] Pad\'{e} approximant to Eq.~(\ref{z2}) then becomes
\begin{equation}
z=2-\frac{\varepsilon}{18}\frac{1+1.790\varepsilon}{1+1.515\varepsilon}\label{z2s}
\end{equation}
with the results for various $d$s given in Table~\ref{exp}. These values are also only slightly different from those that obtained from the two-loop $\eta(d=0)$ \cite{Breuer81}.

As mentioned above, the [3/2] Pad\'{e} approximant to $\eta$ yields the smallest estimates for $d=5$ to 3 and the second smallest for $d=2$ as can be seen from Table~\ref{exp}. We also notice that the direct [2/1] Pad\'{e} approximant to $\eta$ \cite{Alcantara} invariably produces the largest estimates given in Table~\ref{exp}. This suggests us to form a [1/1] Pad\'{e} approximant to $z$, which is
\begin{equation}
z=2-\frac{\varepsilon}{18}\frac{1}{1-0.0153\varepsilon},\label{z11s}
\end{equation}
whose results are indeed all larger than those from Eq.~(\ref{z2s}), although the true values may not necessarily lie in between them. Nevertheless, we take their average as our final estimates for $z$ for $d=5$ to 2 to account for possible bias due to the two-point Pad\'{e} approximant as seen in the estimates of $\eta$.

Knowing $\eta$ and $z$, we can then compute the hysteresis exponents from Eqs.~(\ref{rh}) and (\ref{nhm}) with the results given in Table~\ref{exp}, where we have also shown their spreads due to $\eta$ and/or $z$ as has been done for $\delta$ and $\nu$. Note that we have computed all other exponents from $\eta$ and $z$ rather than resummed their respective $\varepsilon$ series since only these two exponents are independent for the $\varphi^3$ theory.

In Table~\ref{exp}, we have also included the numerical results from direct numerical solutions of Eqs.~(\ref{H}) and (\ref{de}) \cite{zhongl05}. One sees that the agreement between theoretical and numerical results is remarkable. In fact, as pointed out above, even the one-loop results agree fortunately with these results \cite{zhongl05}, though high-order ones without resummations do not, similar to the case of critical phenomena. Of course, higher-order theoretical and further numerical results are desirable. Nevertheless, this agreement confirms again the relevance of the $\varphi^3$ theory to the FOPTs. Note that although we only compare the hysteresis exponents in Table~\ref{exp}, the static exponents along with the dynamic exponent $z$ can also be estimated \cite{Fan11}. As they comprise the hysteresis exponents and are derived directly from the RG theory, they are more fundamental. Note, however, that they have nothing to do with the transition at the usual equilibrium transition point.

\section{\label{discuss}Discussions}
We have studied in detail the $\varphi^3$ theory for the FOPTs. The compelling evidences for scaling in driven FOPTs as mentioned in Sec.~\ref{intro} and demonstrated so far in this paper provides strong evidences for the relevance of the theory and its infrared stable fixed points to the scaling. We have also shown in Sec.~\ref{sol} and in particular through Eq.~(\ref{uks}) that the fixed points are indeed reached independent on the initial coupling when $\kappa\rightarrow0$. However, as the fixed points are imaginary, one may wonder how such imaginary fixed points could be reached from the real physical world, or more specifically, how they could be reached for an RG flow starting from real physical conditions and thus questions the extent of the relevancy. A possible other interpretation is then that the scaling behavior is only a crossover affected by the $\varphi^3$. For the fixed points to affect the flows, however, the latter should flow sufficiently close to the former. Yet, as the fixed points are far away from the real physical parameters at least for practical $\varepsilon$s, it would be hardly possible for them to leave their trace in real measurements. On the other hand, at least in simulations \cite{zhongl05}, scaling is easily found in all spatial dimensions without any detectable indication of difficulty for lower dimensions in which the fixed points are farther off. Moreover, as has been pointed out, the hysteresis exponents found agree well with those of the $\varphi^3$ theory. Therefore, such a crossover is unlikely to be true if not exclusive at all. We also note that the crossover of the pseudo-critical phenomena \cite{Saito} cannot either be a candidate as even in the mean-field level the theory is irrelevant as shown in Sec.~\ref{mfhys}, although the hysteresis exponents that would be predicted by the theory, $n_H=3/5$ for $d\geq4$ from Table~\ref{mfexp} and $n_H=0.5493(12)$ in $d=3$ \cite{zhong06}, are not far away from those listed in Table~\ref{exp}.

\begin{figure}[t]
\centerline{\epsfig{file= 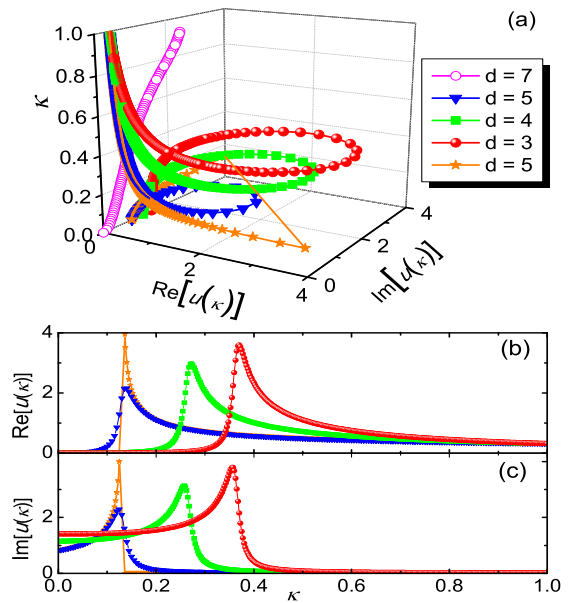,height=8cm,width=13cm}}
\caption{\label{flow}(Color online) RG flows from $\kappa=1$ to $\kappa=0$ of the coupling for a real [$d=5$ (stars)] and a complex (all the others with $d<6$) initial coupling, $u_R^2=0.1$ and $u_R^2=0.1+0.01i$, respectively. The flow for $d=7$ has $u_R^2=0.1+2i$. For clarity of illustration, we have cut off the large values of the flow with a real initial value at $d=5$ (stars). (b) and (c) show the projections of the flows in (a) to the real and imaginary plane, respectively, in the absence of the flow for $d=7$. The legend applies to all panels.}
\end{figure}
In order to see explicitly how the fixed points are reached, we now study the flow equations in detail by regarding them as rigorous equations. Consider the flow of the coupling, Eq.~(\ref{kur}), with $\beta(u_R)$ given by Eq.~(\ref{betau1}). This equation is solved analytically by
\begin{equation}
u_R^2(\kappa)=\frac{2\varepsilon u_R^2}{3u_R^2\left(\kappa^{\varepsilon}-1\right)+2\varepsilon \kappa^{\varepsilon}}\label{urk2}
\end{equation}
for the initial $u_R=u_R(\kappa=1)$. One sees that $u_R^2(\kappa)$ converges to $u_R^{*2}$ and 0 correctly for $\kappa\rightarrow0$ when $\varepsilon$ is positive and negative corresponding to $d<6$ and $d>6$, respectively. The denominator of the solution equals zero at a specific $\kappa_0$ satisfying
\begin{equation}
\kappa_0^{\varepsilon}=\frac{3u_R^2}{3u_R^2+2\varepsilon}\qquad{\rm or,}\qquad\kappa_0={\rm e}^{-{\rm sgn}(\varepsilon)2/3u_R^2}\label{k0}
\end{equation}
to $O(1)$, where ${\rm sgn}$ is the sign function. Because the running variable $\kappa$ and the dimension $\varepsilon$ are real, only real $u_R$ gives rise to a real $\kappa_0$. For $\varepsilon<0$, $\kappa_0>1$. As a result, it does not affect the flow from $\kappa=1$ to $\kappa=0$ and hence any initial $u_R$ will converge continuously to the Gaussian fixed point as expected, as is shown in Fig.~\ref{flow}(a). However, for $\varepsilon>0$, $\kappa_0<1$ and the flow will blow out at $\kappa_0$ for real $u_R$. So, $\kappa_0$ is the branch point of $u_R(\kappa)$. Yet, after approaching infinity, $u_R(\kappa)$ acquires an infinite imaginary part which then converges to the fixed point values as seen in Fig.~\ref{flow}(b) and (c). It seems that the divergence spontaneously triggers an imaginary part in $u_R(\kappa)$ so that the purely imaginary fixed point can still be reached as the solution~(\ref{urk2}) shows. Conversely, one may imagine that in order to acquire an imaginary part to reach the imaginary fixed point, the initial real flow has to enter a divergent ``no-man's" region. Nevertheless, the flow trajectory jumps abruptly from the real plane to the imaginary one as illustrated in Fig.~\ref{flow}(a). However, it becomes continuous once $u_R$ possesses an imaginary part, no matter how small it is. The only difference for different values of the imaginary part is the magnitude of the convolutions shown in Fig.~\ref{flow}(a). The smaller the former is, the larger the latter. Also, no qualitative difference shows between different spatial dimensions below $d_c=6$; the difference is only the value of $\kappa_0$ and again the magnitude of the convolution. We have checked that the divergence remains when considering the contribution from two-loop order, though in that case one has to resort to numerical solutions. As other variables depend on $u_R(\kappa)$, they also exhibit a similar feature.

One sees therefore that if the initial coupling possesses an imaginary value, the fixed points can be continuously reached irrespective of its magnitude, while for purely real initial values to converge to the imaginary fixed points the RG flows have to diverge at a finite scale in order to acquire an imaginary part. At present, we can speculate the reason for the imaginary nature of the fixed points. As noticed in \cite{zhongl05}, it has been shown that the free energy of metastable states is complex \cite{Langer67,Klein83}. In fact, since the $\varphi^3$ theory is not bounded from below, an analytical continuation has to be performed in computing the free energy~(\ref{f}). In particular, as the free energy diverges for negative $\varphi$, the integration in this region has to be deformed to the imaginary axis. So, one should set $\varphi\rightarrow i\varphi$ in the integrand. This naturally leads to $ig_3$. Consequently, an imaginary fixed point appears to be a natural and plausible choice for controlling the local unstable transition near the instability point of an FOPT, just counter possibly to the intuition that only real values are physics. It can also be imagined that the divergence at a finite scale of a real initial coupling should also be pertinent to this. Moreover, note that the negative $\varphi$ is just the direction of the FOPT instead of the third-order transition. However, before a solid solution to the divergence and imaginary problems, the present theory remains a hypothesis, though we have shown clearly its relevance to the FOPTs.

\section{\label{sum}Summary}
We have basically repeated the whole theory of critical phenomena in the $\varphi^3$ theory for FOPTs in an attempt to stress their similarity and differences. Instead of lengthy proofs we have tried to illustrate known general rules with direct examples. We have studied in detail the mean-field theory, the Gaussian theory, the perturbation expansion, and the RG theory. Finite-time scaling and leading corrections to scaling have also been considered. We have also touched on the Yang--Lee edge singularity and employed its results to improve our estimates of the instability exponents. The main results are as follows.

In the mean-field approximation, we have shown that, for a scalar $\phi^4$ model below its $T_c$, the FOPT at the spinodal point driven by an external field is governed by the transition at the instability point of the corresponding $\varphi^3$ model, although the FOPT and the third-order transition between the two states described by the $\varphi^3$ theory fall into opposite ground states and appear different. Via finite-time scaling, numerical results and analytical solutions indeed confirm the relevance of the instability exponents instead of the critical exponents. The Gaussian theory shows clearly that both the correlation length and the correlation time diverge at the instability point similar to the case of the critical phenomena. In addition, all the mean-field instability exponents have been derived conventionally. The perturbation expansion around the mean-field theory then demonstrates as expected that infrared divergences plague and thus necessitate an RG theory for $d\leq6$, while the mean-field results survive for $d>6$ and only outside the unstable region for $d<6$. The power counting analysis shows that in the unstable region and for $d<6$, the effective local field $\varphi^3$ theory does reproduce the sum of the most divergent contributions order by order in the mean-field expansion. Detailed computations of the renormalization functions to one-loop order and derivations and solutions of the RG equations for a general vertex function and the magnetic field then show unambiguously that there is a pair of complex-conjugate imaginary fixed points that are infrared stable similar to the critical fixed point and thus control the large-scale universal behavior. Exact scaling forms and scaling laws among the instability exponents have been derived and the latter have been computed to one-loop order. In six dimensions, these instability exponents recover the mean-field ones, which shows again that the fixed points, albeit imaginary, indeed describes correctly the large-scale fluctuations at the instability point at which the real mean-field transition takes place. We have computed two particular explicit forms of the renormalized massless two-point response function to one-loop order. The finite-time scaling form and associated exponents have also been derived, which serve as an accessible method to probe the instability point. We have derived the leading corrections to scaling and confirmed that they are controlled by a universal exponent.
We have also shown that the infrared behavior of the $\varphi^3$ theory of the FOPTs falls in the same universality class to the Yang--Lee edge singularity though the former is for $T<T_c$ but the latter for $T>T_c$. This implies that the $\varepsilon$ expansions can be trusted and the $\phi^4$ interaction is irrelevant. Moreover, the two- and three-loop-order exponents of the edge singularity have been employed to estimate the instability exponents. The outcomes agree well with previous numerical results, confirming again the relevance of the $\varphi^3$ theory to FOPTs.
It appears that this relevancy is unlikely just a crossover behavior. On the one hand, although the RG flow with a purely real initial coupling diverges at a finite scale, an imaginary part generates beyond that scale and the imaginary fixed point can then be reached. On the other hand, if the initial coupling acquires a finite imaginary part of whatever magnitude, true asymptotic behavior can be surely established. We speculate that the imaginary nature of the fixed points may be a natural consequence of the instability of the $\varphi^3$ theory. However, further studies are clearly needed.

We conclude therefore that the instability fixed points with their instability exponents of the $\varphi^3$ theory are clearly relevant to the scaling and universality behavior exhibited by FOPTs near their instability points. Further studies are desirable to dispel possible concern with the instability point, to clarify the divergence of the RG flow and the imaginary nature of the fixed points, to search for new classes, as well as to find experimental evidences.

\appendix
\section{\label{sa}Supersymmetry action}
Following Ref.~\cite{Justin}, another method to deal with the Jacobian, Eq.~(\ref{jacobi}), is to use the Gaussian integral of anticommuting classical Grassmann variables,
\begin{equation}
\det E=\int {\cal D}c {\cal
D}\tilde{c}\exp\left(\tilde{c}Ec\right),\label{gaussg}
\end{equation}
to write the generating functional as
\begin{equation}
Z[J,\tilde{J}]=\int {\cal D}\phi {\cal D}\tilde{\phi}{\cal D}c
{\cal D}\tilde{c}\exp\left[-{\cal L}+\int dx\left( J\phi
+\tilde{J}\tilde{\phi}\right)\right] \label{zz}
\end{equation}
with the effective action
\begin{eqnarray}
{\cal L}[\phi,\tilde{\phi},c,\tilde{c}] = \int dx\left[\tilde{\phi}\left(
\frac{\partial \phi }{\partial t} + \lambda \frac{\delta {\cal
H}}{\delta \phi }\right)-\lambda \tilde{\phi}^2\right.\nonumber\\
\left.- \tilde{c}\left(\frac{\partial}{\partial
t} + \lambda \frac{\delta^2 {\cal H}}{\delta \phi^2
}\right)c\right], \label{L}
\end{eqnarray}
where
\begin{equation}
c_nc_{n'}+c_{n'}c_n=0,\ \ {\rm for\ any}\ n,n'\label{fermi}
\end{equation}
with $c_n$ representing $c$ or $\tilde{c}$ and use has been made of the local nature of ${\cal H}$.

Introducing two new Grassmann variables $\theta$ and
$\tilde{\theta}$ as additional coordinates to $x$ and defining a superfield $\psi$
by
\begin{equation}
\psi(x,\theta,\tilde{\theta})=\phi(x)+\sqrt{\lambda}\tilde{\theta}
c(x)+\sqrt{\lambda}\tilde{c}(x)\theta+\lambda\tilde{\theta}\theta\tilde{\phi}(x),\label{sup}
\end{equation}
one can write the action (\ref{L}) in a beautiful form
\begin{equation}
{\cal L}[\psi]=\int d\theta d\tilde{\theta}dt
\left\{\frac{1}{\lambda}\int d{\bf x}\tilde{D}\psi D\psi+{\cal
H}[\psi]\right\}\label{spsi}
\end{equation}
with the definitions
\begin{equation}
\tilde{D}\equiv\frac{\partial}{\partial \theta},\ \ D\equiv
\frac{\partial}{\partial
\tilde{\theta}}-\theta\frac{\partial}{\partial t}
\end{equation}
by noting that the integrations over $\theta$ and $\tilde{\theta}$
select the term proportional to $\theta \tilde{\theta}$ due to
their anticommuting character similar to Eq.~(\ref{fermi}). For
example,
\begin{equation}
\int d\theta d\tilde{\theta}dt{\cal
H}[\psi]=\lambda\int dx\left\{\frac{\delta{\cal
H}[\phi]}{\delta\phi(x)}-\tilde{c}(x)\frac{\delta^2{\cal
H}[\phi]}{\delta\phi(x)^2}c(x)\right\}. \nonumber
\end{equation}

The action (\ref{spsi}) is invariant under a supersymmetry transformation
\begin{equation}
\delta\psi=\epsilon\left(\frac{\partial}{\partial\theta}+\tilde{\theta} \frac{\partial}{\partial t}\right)\psi,
\end{equation}
or, in the component form,
\begin{eqnarray}
\delta\phi=\sqrt{\lambda}\tilde{c}\epsilon,\qquad \delta\tilde{c}=0,\nonumber\\
\delta c=\left(\tilde{\phi}-\frac{1}{\sqrt{\lambda}}\frac{\partial\phi}{\partial t}\right)\epsilon, \qquad \delta\tilde{\phi}=\frac{1}{\sqrt{\lambda}}\frac{\partial\tilde{c}}{\partial t}\epsilon,
\end{eqnarray}
which mix commuting and anticommuting fields, where $\epsilon$ here is an infinitesimal anticommuting number. This supersymmetry gives rise to Ward-Takahashi identities which result in the fluctuation-dissipation theorem, Eq.~(\ref{fdt}), when combined with causality \cite{Justin}. It also ensures that the equal-time correlation functions converge at large times to the corresponding static ones \cite{Justin}.

Renormalization can be directly performed to the action (\ref{spsi}) \cite{Justin}. Direct power counting confirms that the static and supersymmetry dynamic theories have the same upper critical dimension. And the supersymmetry then leads to the renormalized action of the form
\begin{equation}
{\cal L}_R[\psi_R]=\int d\theta d\tilde{\theta}dt
\left\{\frac{1}{\lambda_R}Z_{\lambda}\int d{\bf x}\tilde{D}\psi_R D\psi_R+{\cal H}_R[\psi_R]\right\}.\label{rsa}
\end{equation}
So, only one new renormalization factor besides the static ones needs to be introduced. Equation~(\ref{rsa}) also implies that the form of the Langevin equation, Eq.~(\ref{de}), is kept after renormalization.

\section{\label{use}Useful formulas and relevant results}
A useful dimensionally regulated integral is \cite{hooft}
\begin{equation}
\int \frac{d\overline{{\bf k}}}{(\tau+2{\bf p}\cdot{\bf k}+{\bf k}^2)^n}=N_d\frac{\Gamma\left(\frac{d}{2}\right)\Gamma\left(n-\frac{d}{2}\right)}{2\Gamma(n)} (\tau-{\bf p}^2)^{\frac{d}{2}-n},\label{iqn}
\end{equation}
where
\begin{equation}
N_d=\frac{2}{(4\pi)^{\frac{d}{2}}\Gamma\left(\frac{d}{2}\right)}\label{nd}
\end{equation}
is the surface area of a $d$-dimensional sphere divided by $(2\pi)^d$ and the Euler Gamma function $\Gamma(z)$ satisfies
\begin{equation}
\Gamma(z+1)=z\Gamma(z)\label{gz1}
\end{equation}
and has poles at zero and negative integers, as can be seen from its expansion \cite{Ryder}
\begin{equation}
\Gamma(-n+\varepsilon)=\frac{(-1)^n}{n!}\left[\frac{1}{\varepsilon}+\psi_1(n+1)+ O(\varepsilon)\right]\label{gep}
\end{equation}
for zero (note that $0!=1$) and integer $n$ for small $\varepsilon$, where
\begin{equation}
\psi_1(z)=\frac{1}{\Gamma(z)}\frac{d\Gamma(z)}{dz}=-\gamma-\frac{1}{z}+ \sum_{\iota=1}^{n}\left(\frac{1}{\iota} - \frac{1}{z+\iota}\right),\label{psiz}
\end{equation}
which, for the integer $n$, becomes
\begin{equation}
\psi_1(n+1)=-\gamma+\sum_{\iota=1}^{n}\frac{1}{\iota}\label{psin}
\end{equation}
with the Euler constant $\gamma=-\psi_1(1)=0.577$. Using Eqs.~(\ref{gz1}) and (\ref{gep}) at $n=0$, one obtains an expansion
\begin{equation}
\Gamma(1+\varepsilon)=\varepsilon\Gamma(\varepsilon)=1-\gamma\varepsilon+O(\varepsilon^2).\label{g1ep}
\end{equation}
From Eq.~(\ref{psiz}), one also has
\begin{eqnarray}
\Gamma\left(\frac{1}{2}+\varepsilon\right)&=&\Gamma\left(\frac{1}{2}\right)+ \Gamma\left(\frac{1}{2}\right) \psi_1\left(\frac{1}{2}\right)\varepsilon+O(\varepsilon^2)\nonumber\\
&=&\sqrt{\pi}\left[1-(\gamma+2\ln2)\varepsilon\right] +O(\varepsilon^2),
\end{eqnarray}
where $\Gamma(1/2)=\sqrt{\pi}$ and $\psi_1(1/2)=-\gamma-2\ln2$.

A useful formula to calculate graphs is to use the Feynman parameters $x_{\iota}$ to write
\begin{equation}
\frac{1}{\prod_{\iota}^{n} A_{i}^{n_{\iota}}}=\frac{\Gamma(\alpha)}{\prod_{\iota}^n\Gamma(\alpha_{\iota})} \int_0^1\prod_{\iota=1}^{n} \left(dx_{\iota} x_{\iota}^{n_{\iota}-1}\right)\frac{\delta(1-x)}{\left(\sum_{\iota}^n x_{\iota}A_{\iota}\right)^{\alpha}},\label{feyn}
\end{equation}
where $\alpha=\sum\alpha_{\iota}$ and $x=\sum x_{\iota}$.
Using Eqs.~(\ref{feyn}), (\ref{iqn}), and the integral
\begin{equation}
\int_0^1x^{\mu-1}(1-x)^{\nu-1}dx=\frac{\Gamma(\mu)\Gamma(\nu)}{\Gamma(\mu+\nu)},\label{betaf}
\end{equation}
one finds
\begin{widetext}
\begin{eqnarray}
\int d\overline{{\bf k}}\frac{1}{{\bf k}^{2n_1}(\tau+{\bf k}^2)^n}=\frac{1}{2}N_d\tau^{\frac{d}{2}-n-n_1}\frac{\Gamma\left(\frac{d}{2}-n_1\right) \Gamma\left(n+n_1-\frac{d}{2}\right)} {\Gamma(n)},\label{iqqn}\\
\int d\overline{{\bf k}}\frac{1}{{\bf k}^{2n_1}({\bf p}-{\bf k})^{2n_2}}=\frac{1}{2}N_d\left({\bf p}^2\right)^{d/2-n_1-n_2} \frac{\Gamma\left(\frac{d}{2}\right)\Gamma\left(\frac{d}{2}-n_1\right) \Gamma\left(\frac{d}{2}-n_2\right)\Gamma\left(n_1+n_2-\frac{d}{2}\right)} {\Gamma(n_1)\Gamma(n_2)\Gamma(d-n_1-n_2)}.\label{ikpk}
\end{eqnarray}
Again, using Eqs.~(\ref{feyn}) and (\ref{iqn}), one obtains
\begin{eqnarray}
I_1({\bf p})=\int d\overline{{\bf k}}\frac{1}{{\bf k}^{2}\left[{\bf k}^{2}+({\bf p}-{\bf k})^{2}\right]^2}=\frac{1}{2}N_dA\left({\bf p}^2\right)^{d/2-3} \Gamma\left(\frac{d}{2}\right)\Gamma\left(3-\frac{d}{2}\right),\label{i1}\\
I_2({\bf p})=\int d\overline{{\bf k}}\frac{1}{{\bf k}^{2}({\bf p}-{\bf k})^{2}\left[{\bf k}^{2}+({\bf p}-{\bf k})^{2}\right]}=\frac{1}{2}N_dB\left({\bf p}^2\right)^{d/2-3} \Gamma\left(\frac{d}{2}\right)\Gamma\left(3-\frac{d}{2}\right),\label{i2}\\
I_3({\bf p}_1,{\bf p}_2)=\int d\overline{{\bf k}}\frac{1}{{\bf k}^{2}({\bf p}_1-{\bf k})^{2}({\bf p}_2+{\bf k})^{2}}=\frac{1}{2}N_dC\mu^{d-6} \Gamma\left(\frac{d}{2}\right)\Gamma\left(3-\frac{d}{2}\right),\label{i3}
\end{eqnarray}
\end{widetext}
where ${\bf p}_1=\mu{\bf k}_1$, ${\bf p}_2=\mu{\bf k}_2$, and
\begin{eqnarray}
A&=&\int_0^1dx\frac{x^{\frac{d}{2}-2}}{(1+x)^{d-3}},\\
B&=&\int_0^1dx\int_0^{1-x}dy \frac{(x+y)^{\frac{d}{2}-3}(1-y)^{\frac{d}{2}-3}}{(1+x)^{d-3}},\\
C&=&\int_0^1dx\int_0^{1-x}dy \left[x{\bf k}_1+y{\bf k}_2^2-(x{\bf k}_1-y{\bf k}_2)^2\right]^{\frac{d}{2}-3}.\nonumber\\
\end{eqnarray}
For a small $\varepsilon=6-d$, one finds from Eqs.~(\ref{gz1}), (\ref{gep}), and (\ref{g1ep})
\begin{equation}
\Gamma\left(\frac{d}{2}\right)\Gamma\left(3-\frac{d}{2}\right)=\frac{4}{\varepsilon} \left[1-\frac{3}{4}\varepsilon+O(\varepsilon^2)\right],
\end{equation}
which has an $\varepsilon$ pole. Owing to this pole, one can simply set $d=6$ in the integrands of $A$, $B$, and $C$ as only first-order poles appear in the $\varepsilon$ expansion to one-loop order. As a result,
\begin{equation}
A=\frac{1}{8}+O(\varepsilon),\quad B=\frac{1}{4}+O(\varepsilon),\quad C=\frac{1}{2}+O(\varepsilon). \label{abc}
\end{equation}
So,
\begin{eqnarray}
I_1({\bf p})&=&\frac{1}{4\varepsilon}\left[1+O(\varepsilon)\right] N_d\left({\bf p}^2\right)^{d/2-3},\label{i1e}\\
I_2({\bf p})&=&\frac{1}{2\varepsilon}\left[1+O(\varepsilon)\right] N_d\left({\bf p}^2\right)^{d/2-3},\label{i2e}\\
I_3({\bf p}_1,{\bf p}_2)&=&\frac{1}{\varepsilon}\left[1+O(\varepsilon)\right] N_d \mu^{d-6}.\label{i3e}
\end{eqnarray}

\begin{acknowledgments}
This work was supported by the National Natural Science Foundation of China (No.10625420) and the Fundamental Research Funds for the Central Universities.
\end{acknowledgments}

\end{document}